 \documentclass[12pt]{article}

\usepackage{setspace}

\usepackage{amsmath}
\usepackage{graphicx}
\usepackage{natbib}
\usepackage{url} 
\usepackage{latexsym}
\usepackage{amsfonts}
\usepackage{amssymb}
\usepackage{psfrag}
\usepackage{graphicx}
\usepackage{calc, cancel}
\usepackage{graphics, graphicx, caption, subcaption}
\usepackage{latexsym, pgfplots, tikz, url,verbatim,multicol,enumerate,bm,listings,color}
\usepackage{multirow}
\usepackage[justification=centering]{caption}

\usepackage{wrapfig, sidecap,enumitem}
\usepackage{epsfig}
\usepackage{longtable}
\usepackage{booktabs}
\usepackage{hyperref}
\usepackage{cleveref}

\usepackage{moreverb}
\usepackage{xcolor}

\UseRawInputEncoding

\newcommand{\normal}{\mathop{\rm N}}

\newcommand{\Tr}{\mathop{\rm Tr}}
\newcommand{\diag}{\mathop{\rm diag}}
\newcommand{\Lc}{\mathcal{L}}

\newcommand{\doublesum}{\sum_{i=1}^n\sum_{g=1}^G}

\newcommand{\muv}{\mbox{\boldmath{$\mu$}}}

\newcommand{\Sigmav}{\mbox{\boldmath{$\Sigma$}}}

\newcommand{\Thetav}{\mbox{\boldmath{$\Theta$}}}

\newcommand{\etav}{\mbox{\boldmath{$\eta$}}}

\newcommand{\piv}{\mbox{\boldmath{$\pi$}}}

\def\half{\frac{1}{2}}
\def\E{\mathbb E}
\def\R{\mathbb R}

\def\Yv{\mathbf Y}

\def\vv{\mathbf v}
\def\Wv{\mathbf W}
\def\wv{\mathbf w}

\def\yv{\mathbf y}

\def\mv{\mathbf m}
\def\1v{\mathbf 1}
\def\0v{\mathbf 0}

\begin{document}
	\doublespacing
	
	\title{Clustering microbiome data using mixtures of logistic normal multinomial models}
	
	\author{Yuan Fang\footnote{Department of Mathematical Sciences, Binghamton University, State University of New York, 4400 Vestal Parkway East, Binghamton, NY, USA 13902. e: yfang8@binghamton.edu} \and Sanjeena Subedi \footnote{School of Mathematics and Statistics, Carleton University,1125 Colonel By Dr, Ottawa, ON, Canada K1S 5B6 e: sanjeena.dang@carleton.ca}}
	\date{}

	\maketitle

\begin{abstract}
Discrete data such as counts of microbiome taxa resulting from next-generation sequencing are routinely encountered in bioinformatics. Taxa count data in microbiome studies are typically high-dimensional, over-dispersed, and can only reveal relative abundance therefore being treated as compositional. Analyzing compositional data presents many challenges because they are restricted on a simplex. In a logistic normal multinomial model, the relative abundance is mapped from a simplex to a latent variable that exists on the real Euclidean space using the additive log-ratio transformation.  While a logistic normal multinomial approach brings in flexibility for modeling the data, it comes with a heavy computational cost as the parameter estimation typically relies on Bayesian techniques. In this paper, we develop a novel mixture of logistic normal multinomial models for clustering microbiome data. Additionally, we utilize an efficient framework for parameter estimation using variational Gaussian approximations (VGA). Adopting a variational Gaussian approximation for the posterior of the latent variable reduces the computational overhead substantially. The proposed method is illustrated on simulated and real datasets.
\end{abstract}

\textbf{Keywords}:Clustering, Model-based clustering, logistic normal multinomial, Microbiome data, Variational Gaussian approximation

\section{Introduction}
The human microbiome comprises of complex communities of microorganisms including but not limited to bacteria, fungi, and viruses, that inhabit in and on a human body \cite{morgan2012, li2015}. It is estimated that there are approximately $10^{14}$ microbial cells associated with the human body, which is around 10 times the number of human cells \cite{ley2006, fraher2012}. The human microbiome plays a significant role in human health and disease status. There is evidence indicating that microbial dysbiosis may lead to diseases such as cardiovascular diseases \cite{koeth2013}, diabetes \cite{qin2012}, inflammatory bowel disease \cite{greenblum2012}, obesity \cite{turnbaugh2009}, and many others. 
Next generation sequencing techniques, such as the 16S ribosomal RNA (rRNA) amplicon sequencing or shotgun metagenomics sequencing, provide an effective way for quantification and comparison of the bacterial composition, including types and abundance of different bacteria within biological samples \cite{streit2004, kuczynski2012, yatsunenko2012, aijo2018}. 
In 16S rRNA sequencing, the 16S rRNA, which is ubiquitous in all bacterial organisms but it also has distinct variable regions that can be used to discriminate between different bacteria is first PCR-amplified and then sequenced \cite{kuczynski2012}. Shotgun sequencing on the other hand is an untargeted sequencing of all microbial genomes in a sample \cite{quince2017}. In either case, short reads are preprocessed through steps of quality control and filtering steps. The processed raw sequence reads are then clustered into operational taxonomic units (OTUs) at a certain similarity level \cite{eckburg2005} where each OTU is characterized by a representative DNA sequence that could be assigned to a taxonomic lineage by comparing to a known database.\cite{li2015} Resulting read counts at different taxonomic levels for $n$ samples over $K+1$ taxa are stored as a $n\times (K+1)$ matrix $\mathbf{W}$, with the entry $W[i,k]$ representing the counts recorded for the $k^{th}$ taxon in the $i^{th}$ sample.\par

Statistical analysis of microbiome data is complicated. The microbiome count data can only reveal relative abundance, i.e., the abundance for each taxa are constrained by the total sum of the microbes in that particular sample and the total sum of microbes could vary among the samples depending on the sequencing depth. Different individuals could share various communities of microorganisms, with only a few major ones in common, and even for one person, the microbial composition could be totally different in different body sites. The heterogeneity of the microbiome samples also leads to over-dispersion. See Hamady and Knight \cite{hamady2009} for a detailed review of challenges related to analyzing microbiome data. Standard multivariate analysis usually fails to capture these properties of the microbiome data. Different models have been proposed for the microbiome counts in the literature that capture one or more of the above intrinsic characteristics such as the negative binomial model\citep{zhang2017}, zero-inflated negative  binomial model\cite{zhang2020}, zero-inflated Poisson model\cite{joseph2013,xu2020}, Dirichlet-multinomial model \citep{holmes2012,chen2013,subedi2020}, and the logistic normal multinomial model\cite{xia2013}. While modelling such count data, a negative binomial (NB) model can allow for the variance to be larger than the mean using a dispersion parameter, thus handling over-dispersion better than a simple Poison model. The zero-inflated negative  binomial (ZINB) and zero-inflated Poisson (ZIP) have been proposed to account for excessive number of zeros \cite{joseph2013}. Xu et al \cite{xu2015} provides a comparison among the zero-inflated models. However, the NB and ZINB model ignore the compositional nature of these microbial counts. Chen and Li \cite{chen2013}, Holmes et al \cite{holmes2012}, Wadsworth et al \cite{wadsworth2017} and Subedi et al \cite{subedi2020} utilized the Dirichlet-multinomial model for microbial counts that takes into account the compositional nature of these data. Alternately, Xia et al \cite{xia2013} employed the logistic normal multinomial model, mapping the relative abundance from a simplex to a latent variable that exists on the real Euclidian space using the additive log-ratio transformation. Cao et al \cite{cao2017} exploited a Poisson-multinomial model and performed a multi-sample estimation of microbial composition in positive simplex space from a high-dimensional sparse count table. Caporaso et al \cite{caporaso2011} quantified variations of microbial composition across time by projecting the dynamics using low-dimensional embedding. {\"A}ij{\"o} et al\cite{aijo2018} proposed a temporal probabilistic model for the microbiome composition using a hierarchical multinomial model. Silverman et al\cite{silverman2018} also developed a dynamic linear model based on the logistic normal multinomial model to study the artificial human guts microbiome.\par

Clustering microbiome samples into groups that share similar microbial compositional patterns is of great interest \cite{holmes2012}. Clustering algorithms are usually categorized into hierarchical clustering and distance-based clustering. Hierarchical clustering has been applied for clustering microbiome data, yet it requires the choice of a cut-off threshold, according to which samples can be divided into groups \cite{holmes2012}. On the other hand, $k-$means clustering, a distance-based method, might not be appropriate for microbiome compositions because it is typically used for continuous data and obtains spherical clusters. Hence, model-based clustering approaches that utilize a finite mixture model have been widely used in the last decade to cluster microbiome data \citep{holmes2012,subedi2020}. A finite mixture model assumes that the population consists of a finite mixture of subpopulations (or clusters), each represented by a known distribution \cite{mclachlan2000, zhong2003, fruhwirth2006, mcnicholas2016}. Due to the flexibility in choosing component distributions to model different type of data, several mixture models based on discrete distributions have been developed to study count data, especially, for gene expression data. Rau et al \cite{rau2011} proposed a clustering approach for RNA-seq data using mixtures of univariate Poisson distributions, Papastamoulis et al \cite{papastamoulis2016} proposed a mixture of Poisson regression models; Si et al \cite{si2014} studied model-based clustering for RNA-seq data using a mixture of negative binomial (NB) distributions; Silva et al \cite{silva2019} proposed a multivariate Poisson-log normal mixture model for clustering gene expression data. However, due to the compositional nature of microbiome data, none of the about discrete mixture models can be employed directly for clustering microbiome data. Holmes et al \cite{holmes2012} adopted the Dirichlet-multinomial (DM) model, where the underlying compositions are modeled as a Dirichlet prior to a multinomial distribution that describes the taxa counts, and proposed a mixture of DM models to cluster samples. \par 

In this paper, we develop a model-based clustering approach using the logistic normal multinomial model proposed by Xia et al \cite{xia2013} to cluster microbiome data. In the logistic normal multinomial model, the observed counts are modeled using a multinomial distribution, and the relative abundance is regarded a random vector on a simplex, which is further mapped to a latent variable that exists on the real Euclidean space through an additive log-ratio transformation. While this approach captures the additional variability compared to a multinomial model, it  does not possess a closed form expression of the log-likelihood functions and of the posterior distributions of the latent variables. Therefore, the expected complete-data log-likelihoods needed in the E-step of a traditional EM algorithm are usually intractable. In such a scenario, one commonly used approach is a variant of the EM algorithm that relies on Bayesian techniques using Markov chain Monte Carlo (MCMC); however, this would typically bring in high computational cost. Here, we develop a variant of the EM algorithm, here on referred to as a variational EM algorithm for parameter estimation that utilizes variational Gaussian approximations (VGA). In Variational Gaussian approximations (VGA)\cite{barber1998}, a complex posterior distribution is approximated using computationally convenient Gaussian densities by minimizing the Kullback-Leibler (KL) divergence between the true and the approximating densities \cite{bishop2006,arridge2018}. Adopting a variational Gaussian approximation delivers accurate approximations of the complex posterior while reducing computational overhead substantially. Hence, this approach has become extremely popular in many different fields in machine learning. \cite{barber1998, bishop2006, archambeau2007, khan2012, challis2013, blei2017}.

The contribution of the paper is two folds - first, we develop a computationally efficient framework for parameter estimation for logistic normal multinomial model through the use of variational Gaussian approximations and second, we utilize this framework to develop a model-based clustering framework for clustering microbiome data. Through simulations and applications to microbiome data, the utilities of the proposed approach is illustrated. The paper is structured as follows: Section \ref{datamodel} describes the logistic normal multinomial model for microbiome count data and details the variational Gaussian approximations. Section \ref{mixture_fmm} provides a mixture model framework based on the model described in Section \ref{datamodel} together with a variational EM algorithm for parameter estimation. In Section \ref{results}, clustering results are illustrated by applying the proposed algorithm on both simulated and real data. Finally, discussion on the advantages and limitations along with some future directions are provided in Section \ref{discussion}

\section{Methodology}\label{datamodel}
\subsection{The logistic normal multinomial model for microbiome compositional data}\label{model_lnmtrans}
Suppose we have $K+1$ bacterial taxa for a sample denoted as a random vector $\Wv = (W_1,\dots,W_{K+1})^\top$. Here, the taxa could represent any level of the bacterial phylogeny such as OTU, species, genus, phylum, etc. Due to the fact that taxa counts from 16S sequencing can only reveal relative abundance, let's suppose there is a vector $\Thetav = (\Theta_,\dots,\Theta_{K+1})$ such that $\sum_{k=1}^{K+1}{\Theta_k} = 1$, which represents the underlying composition of the bacterial taxa. Then, the microbial taxa count $\Wv $ can be modeled as a multinomial random variable with the following conditional density function:
\begin{equation*}
p(\wv|\Thetav) \propto \prod_{k=1}^{K+1}(\Theta_k)^{w_k}.
\end{equation*}
Several models have been proposed in the literature that capture the relative abundance nature of microbiome data and analyze the compositional data \cite{holmes2012,xia2013}. Here we use the model by Xia et al \cite{xia2013} that utilizes an additive log-ratio transformation $\phi(\Thetav)$ proposed by Aitchison \cite{aitchison1982} such that:
\begin{equation}\label{alr}
\Yv = \phi(\Thetav) = \left(\log\left(\dfrac{\Theta_1}{\Theta_{K+1}}\right),\dots,\log\left(\dfrac{\Theta_K}{\Theta_{K+1}}\right)\right)^\top.
\end{equation}
This transformation $\phi$ maps the vector $\Thetav$ from a $K$-dimensional simplex to the $K$-dimensional real space $\R^K$ while $\Yv$ is assumed to follow a multivariate normal distribution with mean $\muv$ and covariance $\Sigmav$ with the density function
\begin{equation*}
p(\yv|\muv,\Sigmav) \propto |\Sigmav|^{-\half}\exp\left\{-\half(\yv-\muv)^\top\Sigmav^{-1}(\yv-\muv)\right\}.
\end{equation*}
As this additive log-ratio transformation is a one-to-one map, the inverse operator of $\phi$ exists and is given by
\begin{equation*}
\Thetav = \phi^{-1}(\Yv)=\begin{cases} 
\dfrac{\exp(Y_k)}{1+\sum_{k=1}^{K}{\exp(Y_k)}} & k=1,\dots,K \\
\dfrac{1}{1+\sum_{k=1}^{K}{\exp(Y_k)}} & k= K+1
\end{cases}.
\end{equation*}
Hence, the joint density of $\Wv$ and $\Yv$ up to a constant is as follows:
\begin{equation*}
\begin{split}
p(\wv,\yv) & \propto p\left(\wv|\phi^{-1}(\yv)\right)p(\yv|\muv,\Sigmav)=\prod_{k=1}^{K+1}\left({\phi^{-1}(\yv)}_k\right)^{w_k}\times |\Sigmav|^{-\half}\exp\left\{-\half(\yv-\muv)^\top\Sigmav^{-1}(\yv-\muv)\right\}
\end{split}.
\end{equation*}
\subsection{A variational Gaussian lower bound}\label{model_vga}
For the microbiome data, only the count vector $\Wv$ are observed while the latent variable $\Yv$ is unobserved. The marginal density of $\Wv$ can be written as
\begin{equation*}
p(\wv) = \int_{\R^K} p(\wv,\yv) d\yv \propto \int_{\R^K} \prod_{k=1}^{K+1}\left({\phi^{-1}(\yv)}_k\right)^{w_k}\times |\Sigmav|^{-\half}\exp\left\{-\half(\yv-\muv)^\top\Sigmav^{-1}(\yv-\muv)\right\} d\yv.
\end{equation*} 
Note that this marginal distribution of $\Wv$ involves multiple integrals and cannot be further simplified. Here, in presence of missing data, an expectation-maximization algorithm \cite{dempster1977} or some variant of it is typically utilized for parameter estimation. An EM-algorithm comprises two steps: an E-step in which the expected value of the complete data (i.e. observed and missing data) log-likelihood is computed given the observed data and current parameter estimate and an M-step in which the complete data log-likelihood is maximized. These step are repeated until convergence to obtain the maximum likelihood estimate of the parameters. To compute the expected value of the complete data log-likelihood, $\E(\Yv\mid\wv)$ and $\E(\Yv\Yv^T\mid\wv)$ needs to be computed for which we need $p(\yv|\wv)$. Mathematically,
\begin{align*}
p(\yv|\wv)=\frac{p(\wv,\yv)}{p(\wv)} = \frac{\prod_{k=1}^{K+1}\left({\phi^{-1}(\yv)}_k\right)^{w_k}\times |\Sigmav|^{-\half}\exp\left\{-\half(\yv-\muv)^\top\Sigmav^{-1}(\yv-\muv)\right\}}{\int_{\R^K}\prod_{k=1}^{K+1}\left({\phi^{-1}(\yv)}_k\right)^{w_k}\times |\Sigmav|^{-\half}\exp\left\{-\half(\yv-\muv)^\top\Sigmav^{-1}(\yv-\muv)\right\}d\yv}.
\end{align*}
However, the denominator involves multiple integrals and cannot be further simplified. One could employ a Markov chain Monte Carlo (MCMC) approach to explore the posterior state space; however, these methods are typically computational expensive, especially for high-dimensional problems.  Here, we propose the use of variational Gaussian approximation (VGA)  \cite{barber1998}  for parameter estimation. A VGA aims to find an optimal and tractable approximation that has a Gaussian parametric form to approximate the true complex posterior by minimizing the Kullback-Leibler divergence between the true and the approximating densities. It has been successfully used in many practical applications to overcome this challenge.\cite{bishop2006, archambeau2007,wainwright2008, khan2012, challis2013, arridge2018}. In order to utilize VGA, we define a new latent variable $\etav$ by transforming $\Yv$ such that
\begin{equation}\label{trans}
\etav = B\Yv, \quad \text{ where }B=\begin{pmatrix}
1&0&\dots&0\\
0&1&\dots&0\\
\vdots&\vdots&\dots&\vdots\\
0&0&\dots&1\\
0&0&\dots&0
\end{pmatrix},
\end{equation}
is a $(K+1)\times K$ matrix which takes the form as an identity matrix attached by a row of K zeros. Given that $\Yv\sim\normal(\muv,\Sigmav)$, the new latent variable $\etav\sim \normal(\tilde{\muv},\tilde{\Sigmav})$ where
\begin{equation}\label{musigtrans}
\tilde{\muv} = B\muv = (\muv,0)^\top; \quad \tilde{\Sigmav} = B\Sigmav B^\top = \begin{pmatrix}
\Sigmav&\vline &\0v_{K\times 1}\\
\hline
\0v_{1\times K}&\vline&0
\end{pmatrix}.
\end{equation} Then, the underlying composition variable $\Thetav$ can be written as a function of $\etav$:
\begin{equation}\label{inv_alr}
\Thetav = \tilde{\phi}^{-1}(\etav) = \dfrac{\exp{\eta_k}}{\sum_{k=1}^{K+1}{\exp{\eta_k}}} k = 1\dots,K+1.
\end{equation}
Suppose we have an approximating density $q(\etav)$, then the marginal log density of $\Wv$ can be written as:
\begin{align*}
\log p(\wv)&= \int \log p(\wv) ~q(\etav) ~d\etav\\
&=\int \log \frac{p(\wv,\etav)/q(\etav)}{p(\etav\mid\wv)/q(\etav)}~q(\etav)~d\etav\\
&= \int \left[\log~p(\wv,\etav) -\log q(\etav)\right]~q(\etav)~d\etav+\int \log\dfrac{q(\etav)}{p(\etav|\wv)}~ q(\etav)~ d\etav \\
&= F(q(\etav),\wv) +D_{KL}(q||p),
\end{align*}
where the first part $F(q(\etav),\wv) = \int q(\etav)\log\dfrac{p(\wv,\etav)}{q(\etav)} d\etav$ is called the evidence lower bound (ELBO) \cite{barber1998} and the second part $D_{KL}(q||p)= \int \log\dfrac{q(\etav)}{p(\etav|\wv)}~ q(\etav)~d\etav$ is the Kullback-Leibler divergence from $p(\etav|\wv)$ to $q(\etav)$. Hence, minimizing the Kullback-Leibler divergence is equivalent to maximizing the following evidence lower bound (ELBO). In VGA, we assume $q(\etav)$ is a Gaussian distribution, such that
\begin{equation*}
q(\etav) = \normal(\etav|\mv,V) \propto |V|^{-\half}\exp\left\{-\half(\etav-\mv)^\top V^{-1}(\etav-\mv)\right\}.
\end{equation*}
Given the fact that $q(\etav)$ is fully characterized by its mean vector and covariance matrix, the above lower bound is a function of the variational parameters $\mv$ and $V$ and we aim to find the optimal set of $(\mv,V)$ such that it maximizes $F(q(\etav,\wv))$. $F(q(\etav,\wv))$ can be separated into three parts:
\begin{equation*}
F(q(\etav),\wv) = F(\mv,V) = - \int q(\etav)\log q(\etav) d\etav + \int q(\etav)\log p(\etav) d\etav + \int q(\etav)\log p(\wv|\etav) d\etav.
\end{equation*}\par

Up to a constant, the last integral, which is denoted as $\gamma$, in the above decomposition is given as follows:
\begin{equation*}
\begin{split}
\gamma &= \int q(\etav)\log p(\wv|\etav) d\etav\\
&= \E_{q(\etav|\mv,V)}\left[\wv^\top\etav-\sum_{k=1}^{K+1}{w_k\log\left(\sum_{k=1}^{K+1}{\exp\eta_k}\right)}\right]\\
&= \wv^\top\mv - \left(\sum_{k=1}^{K+1} w_k\right)\E_{q(\etav|\mv,V)}\left[\log\left(\sum_{k=1}^{K+1}{\exp\eta_k}\right)\right].
\end{split}
\end{equation*}
Similar to Blei and Lafferty,\cite{blei2006} we use an upper bound for the expectation of log sum exponential term with a Taylor expansion,
\begin{equation*}
\E_{q(\etav|\mv,V)}\left[\log\left(\sum_{k=1}^{K+1}{\exp\eta_k}\right)\right] \leq \xi^{-1}\left\{\sum_{k=1}^{K+1}\E_{q(\etav|\mv,V)}\left[\exp(\eta_k)\right]\right\}-1+\log(\xi),
\end{equation*}
where $\xi\in \R$ is introduced as a new variational parameter.\par
Here, we further assume that $V$ is a diagonal matrix with the first $K$ diagonal element of $V$ as $v_k^2$  and the $K+1^{th}$ diagonal element is set to 0 such that
\begin{equation*}
v_k^2 = \begin{cases}
v_k^2,&k=1,\dots,K\\
0,&k=K+1.
\end{cases}
\end{equation*}
We also denote the $k-$th element of $\mv$ as $m_k$ such that
\begin{equation*}
m_k = \begin{cases}
m_k, &k=1,\dots,K\\
0,&k=K+1.
\end{cases}
\end{equation*}
Hence, the expectation 
\begin{equation*}
\E_{q(\etav|\mv,V)}\left[\exp(\eta_k)\right]=
\exp\left(m_k+\dfrac{ v_k^2}{2}\right), \text{ for }k=1,\dots,K+1. 
\end{equation*}
Based on this upper bound, we obtain a concave lower bound to $\gamma$ and to the ELBO. The new concave variational Gaussian lower bound to the model evidence $\log p(\wv)$ is given as follows
\begin{equation}\label{vglb}
\begin{split}
\tilde{F}\left(\mv,V,\tilde{\muv},\tilde{\Sigmav},\xi\right) &= \wv^\top\mv - \left(\sum_{k=1}^{K+1} w_k\right)\left\{\xi^{-1}\left[\sum_{k=1}^{K+1}\exp\left( m_k+\dfrac{ v_k^2}{2}\right)\right]-1+\log(\xi)\right\}\\
&-\half\log|B^\top\tilde\Sigmav B|-\half(\mv-\tilde \muv)^\top\tilde\Sigmav^{\ast}( \mv-\tilde \muv)-\half \Tr(\tilde\Sigmav^{\ast} V)\\
&+\half\sum_{k=1}^K\log( v_k^2)+\dfrac{K}{2},
\end{split}
\end{equation}
where $$\tilde\Sigmav^{\ast} = \begin{pmatrix}
\Sigmav^{-1}&\vline &\0v_{K\times 1}\\
\hline
\0v_{1\times K}&\vline&0
\end{pmatrix},$$ is the generalized inverse of $\tilde{\Sigmav}$.
Details on the derivation of this lower bound can be found in Appendix~\ref{math_detail}. Given fixed $\wv$, $\tilde \muv$, and $\tilde \Sigmav$, this lower bound only depends on the variational parameter set $(\mv,V,\xi)$.\par  
Maximization of the lower bound $\tilde{F}\left(\mv,V,\tilde{\muv},\tilde{\Sigmav},\xi\right)$ with respect to $\xi$ has a closed form solution and is given by 
\begin{equation}\label{xiop}
\hat{\xi} = \sum_{k=1}^{K+1}\exp\left(m_k+\dfrac{v_k^2}{2}\right).
\end{equation}
However, maximization with respect to $ \mv$ and $v_k, k=1,\dots,K$ do not possess analytical solutions. We use Newton's method to search for roots to the following derivatives:
\begin{equation}\label{Fprimem}
\dfrac{\partial \tilde{F}}{\partial \mv} = \wv - \tilde{\Sigmav}^{\ast}( \mv-\tilde \muv)-\left(\sum_{k=1}^{K+1} w_k\right)\xi^{-1}\exp\left(\mv+\dfrac{\vv^2}{2}\right),
\end{equation}
with $\vv^2=(v_1^2,\dots,v_K^2,0)$ denoting the diagonal element of $V$ as a vector; and 
\begin{equation}\label{Fprimev}
\dfrac{\partial \tilde{F}}{\partial  v_k} =  v_k^{-1}- v_k\tilde \Sigmav_{k,k}^{\ast} - \left(\sum_{k=1}^{K+1} w_k\right)\xi^{-1}\exp\left( m_k+\dfrac{ v_k^2}{2}\right) v_k.
\end{equation}
Details can be found in Appendix~\ref{math_detail}.

\subsection{Mixture of logistic normal multinomial models}\label{mixture_fmm}
Assume there are $G$ subgroups in the population, with $\pi_g$ denoting the mixing weight of the $g-$th component such that $\sum_{g=1}^{G}\pi_g = 1$. Then, a $G-$component finite mixtures logistic normal multinomial models can be written as
\begin{equation*}
f(\wv\mid \boldsymbol{\vartheta})= \sum_{g=1}^G \pi_g f_g(\wv\mid\vartheta_g),
\end{equation*}
where $f_g(\wv\mid\boldsymbol{\vartheta}_g)$ represents the density function of the observation $\Wv = \wv$, given that $\Wv$ comes from the $g-$th component with parameters $\vartheta_g$.\par 
Provided $n$ observed counts, $\wv = (\wv_1,\dots,\wv_n)$ with a transformed underlying the composition $\Yv_i, i= 1,\dots, n$, the likelihood of a $G-$component finite mixture is given as
\begin{equation*}
\Lc(\boldsymbol{\vartheta}\mid\wv)=\prod_{i=1}^{n}f(\mathbf{w}_i\mid \boldsymbol{\vartheta})= \prod_{i=1}^{n}\sum_{g=1}^G \pi_g f_g(\mathbf{w}_i\mid\vartheta_g) = \prod_{i=1}^{n}\sum_{g=1}^G \pi_g \int p(\wv_i\mid\yv_i)p(\yv_i\mid\vartheta_g) d\yv_i.
\end{equation*}
In clustering, the unobserved component membership is denoted by an indicator variable $z_{ig}, i=1,\dots,n, g=1,\dots,G$ that takes the form \begin{equation*}
z_{ig} = \begin{cases}
1 & \text{ if the } i-th \text{ observation is from the } g-th \text{ group},\\
0 & \text{ otherwise}.
\end{cases}
\end{equation*}

Therefore, conditional on $z_{ig}$, we have 
\begin{equation*}
\Yv_i|z_{ig} = 1 \sim \normal(\muv_g,\Sigmav_g).
\end{equation*}
In order to utilize the variational approach for parameter estimation, we again define a new latent variable $\etav$ such that $\etav = B\Yv$ and
$$\etav_i\mid z_{ig}=1 \sim \normal(\tilde{\muv}_g,\tilde{\Sigmav}_g), ~\text{where}~
\tilde{\muv}_g = B\muv_g = (\muv_g,0)^\top~ \text{and}~ \tilde{\Sigmav}_g = B\Sigmav_g B^\top = \begin{pmatrix}
\Sigmav_g&\vline &\0v_{K\times 1}\\
\hline
\0v_{1\times K}&\vline&0
\end{pmatrix}.
$$
Therefore, the complete data (i.e., observed counts $\Wv$ and unobserved class label indicator variable) log-likelihood using the marginal density of $\Wv$ is
\begin{align*}
\ell &=\log \left[\prod_{i=1}^{n}\prod_{g=1}^G \pi_g f_g(\mathbf{w}_i\mid\vartheta_g)\right]^{z_{ig}}=\doublesum z_{ig}\left\{\log \pi_g + \log\left[\int p(\wv_i|\etav_i)p(\etav_i|\tilde \muv_g,\tilde \Sigmav_g)d\etav_i\right]\right\}.
\end{align*}
To perform variational inference on the mixture model, we substitute $\log\left[\int p(\wv_i|\etav_i)p(\etav_i|\muv_g,\Sigmav_g) d\etav_i \right]$ by the variational Gaussian lower bound $\tilde{F}\left(\mv,V,\tilde{\muv},\tilde{\Sigmav},\xi\right)$ derived in Section~\ref{model_vga}. Hence, the variational Gaussian lower bound of complete data log likelihood can be written as:
\begin{equation}\label{vgcomplik}
\begin{split}
\tilde\Lc = &\doublesum z_{ig}\log \pi_g + \doublesum z_{ig} \wv_i^\top\mv_{ig}\\
&-\doublesum z_{ig}\left(\sum_{k=1}^{K+1} w_{ik}\right)\left\{\xi_i^{-1}\left[\sum_{k=1}^{K+1}\exp\left( m_{igk}+\dfrac{ v_{igk}^2}{2}\right)\right]-1+\log(\xi_i)\right\}\\
&+\doublesum z_{ig}\left\{\half\log|B^\top\tilde\Sigmav_g B|-\half(\mv_{ig}-\tilde \muv_g)^\top\tilde\Sigmav_g^{\ast}(\mv_{ig}-\tilde\muv_g)\right.\\
&\quad \quad \quad \quad \quad \quad \quad \left.-\half\Tr(\tilde\Sigmav_g^{\ast} V_{ig})+\half\sum_{k=1}^K\log( v_{igk}^2)+\dfrac{K}{2}\right\}.
\end{split}
\end{equation}
Hence, we need to find optimal solutions to variational parameters $(\mv_{ig},V_{ig},\xi_i)$ that are associated with each observation $\wv_i, i=1,\dots,n$, as well as the model group-specific Gaussian parameters $(\tilde{\muv}_g, \tilde{\Sigmav}_g), g=1,\ldots, G$, such that the complete data variational Gaussian lower bound $\tilde\Lc$ is maximized. The use of VGA provides great reduction in the computational time. 

\subsection{The variational EM algorithm}
Parameter estimation can be done in an iterative EM-type approach, from here on referred to as variational EM such that the following steps are iterated until convergence. For the parameters that do not have a closed form solution to the optimization, we perform one step of Newton's method to approximate the root to their first derivatives.
\begin{enumerate}
\item[Step 1:] Conditional on the variational parameters $(\mv_{ig},V_{ig},\xi_i)$ and model group-specific Gaussian parameters $(\tilde{\muv}_g, \tilde{\Sigmav}_g)$, $\E(Z_{ig}\Wv_i)$ is computed.
Given $(\tilde{\muv}_g, \tilde{\Sigmav}_g)$,
\begin{equation*}
\E\left(Z_{ig}\mid \wv_i\right)=\frac{\pi_g f_g(\wv_i\mid\tilde \muv_g,\tilde \Sigmav_g)}{\sum_{h=1}^G\pi_h f_h(\wv_i\mid\tilde \muv_h,\tilde \Sigmav_h)}.
\end{equation*}
This involves the marginal distribution of $\Wv$ and hence, we use an approximation of $\E\left(Z_{ig}\mid \wv_i\right)$ where we replace the marginal density $\Wv$  by the exponent of ELBO such that
\begin{equation*}
\hat{z}_{ig}:=\dfrac{\pi_g \exp\left\{ \tilde{F}\left(\wv_i,\mv_{ig},V_{ig},\tilde{\muv}_{g},\tilde{\Sigmav}_{g},\xi_i\right)\right\}}{\sum_{j=1}^{G}\pi_j \exp\left\{ \tilde{F}\left(\wv_i,\mv_{ij},V_{ij},\tilde{\muv}_{j},\tilde{\Sigmav}_{j},\xi_i\right)\right\}}.
\end{equation*}
\item[Step 2:] Update $\hat{\xi}_i,\hat{\mv}_{ig},\hat{V}_{ig}$: 
		\begin{itemize}
			\item update $\hat{\xi}_i$ according to Equation~\ref{xiop};
			\item update $\hat{\mv}_{ig}$ by performing one step of Newton's method for approximating the root to the derivative in Equation~\ref{Fprimem}, then let $\hat{m}_{ig(K+1)}=0$;
			\item for $k=1,\dots,K$, update $\hat{v}_{igk}^2$ by performing one step of Newton's method searching root to the derivative in Equation ~\ref{Fprimev}, let $\hat{v}_{ig(K+1)}^2=0$, then $\hat{V}_{ig} = \diag(\hat{v}_{ig1}^2,\dots,\hat{v}_{ig(K+1)}^2)$.
		\end{itemize}
		
\item[Step 3:]  Update $\pi_{ig}$, $\tilde\muv_g$ and $\tilde\Sigmav_g$ as
\begin{align*}
\hat{\pi}_{ig}&=\frac{\sum_{n=1}^n \hat{z}_{ig}}{n},\\
\hat{\tilde{\muv}}_g &= \dfrac{\sum_{i=1}^n \hat{z}_{ig}\hat{\mv}_{ig}}{\sum_{i=1}^n \hat{z}_{ig}},\\
\hat{\tilde{\Sigmav}}_g &= \dfrac{\sum_{i=1}^n \hat{z}_{ig} \left[\hat{V}_{ig}+(\hat{\mv}_{ig}-\hat{\tilde{\muv}}_g)(\hat{\mv}_{ig}-\hat{\tilde{\muv}}_g)^\top\right]}{\sum_{i=1}^n \hat{z}_{ig}}.
\end{align*}
\end{enumerate}
Note that the original parameters $\muv_{g}$ and $\Sigmav_{g}$ can be obtained by the transformation
		\begin{equation*}
		\hat\muv_{g} = B^\top\hat{\tilde{\muv}}_g;\quad  \hat\Sigmav_{g} = B^\top \hat{\tilde{\Sigmav}}_g B.
		\end{equation*}

An Aitken acceleration criterion\cite{aitken26}  is employed to stop the iterations. More specifically, at $t^{th}$ iteration, when $t >2$, calculate
	\begin{equation*}
	\begin{split}
	a^{(t-1)} &= \dfrac{\ell^{(t)}-\ell^{(m-1)}}{\ell^{(t-1)}-\ell^{(t-2)}},\\
	\ell_{\infty}^{(t)} &= \ell^{(t-1)}+\dfrac{1}{1-a^{(t-1)}}\left(\ell^{(t)}-\ell^{(t-2)}\right),
	\end{split}
	\end{equation*}
	where $\ell^{(t)}=\tilde{F}\left(\wv_i,\mv_{ig},V_{ig},\tilde{\muv}_{g},\tilde{\Sigmav}_{g},\xi_i\right)$ is the variational Gaussian lower bound who approximates the log likelihood at $t^{th}$ iteration. Then, the algorithm will be stopped when $\left\vert\ell_{\infty}^{(t)}-\ell_{\infty}^{(t-1)}\right\vert < \epsilon$ for a given $\epsilon$\cite{bohning1994}.  In our analysis, we $\epsilon$ is set to be $1\times10^{-3}$. 

\subsubsection{Hybrid Approach}
While the VGA based approach only approximates the posterior distribution and it does not guarantee exact posterior \citep{ghahramani1999}, it is computationally efficient. On the other hand, a fully Bayesian MCMC based approach can generate exact results, fitting such models can take substantial computational time. For example, fitting one iteration using a fully Bayesian MCMC model for a five dimensional dataset (from Simulation study 1) with $n=1000$ takes on average of 45 minutes. In a clustering context, the number of iterations required for the analysis is typically in hundreds. 
 Thus, we provide a computationally efficient hybrid approach in which 
\begin{itemize}[topsep=0pt,itemsep=-3pt]
\item[--] Step 1: Fit the model using the VGA based approach.
\item[--] Step 2: Estimate the component indicator variable $Z_{ig}$ conditional on the parameter estimates from the VGA based approach.
\item[--] Step 3: Using the parameter estimates from Step 1 as the initial values for the parameters and using the classification from Step 2, compute the MCMC based expectation for the latent variable $\tilde\etav_{ig}$ as:
\begin{equation*}
\begin{split}
\label{eqn:Ezig_chapter5}
& \E ( \tilde{\etav}_{ig} | \mathbf{W}_i) \simeq \frac{1}{R} \sum_{k=1}^R \tilde{\etav}_{ig}^{(k)}.
\end{split}
\end{equation*}
and $ \boldsymbol{\theta}_{ng}^{(k)}$ is a random sample from the posterior distribution of  $\tilde\etav_{ig}$ simulated via the ${\tt RStan}$ package for iterations $k= 1,\ldots, R$ (after discarding the burn-in). 
\item[--] Step 4: Obtain the final estimates of the model parameters as:
\begin{align*}
\hat{\pi}_{ig}&=\frac{\sum_{n=1}^n \hat{z}_{ig}}{n},\\
\hat{\tilde{\muv}}_g &= \frac{\sum_{n=1}^n \hat{z}_{ig}\E\big(\tilde{\etav}_{ig}\big)} {\sum_{n=1}^n \hat{z}_{ig}},\\
\hat{\tilde{\Sigmav}}_g &= \dfrac{\sum_{i=1}^n \hat{z}_{ig} \E\left[(\hat{\tilde\etav}_{ig}-\hat{\tilde\muv}_g)(\hat{\tilde\etav}_{ig}-\hat{\tilde{\muv}}_g)^\top\right]}{\sum_{i=1}^n \hat{z}_{ig}}.
\end{align*}
\end{itemize}
The hybrid approach comes with a substantial reduction in computational overhead compared to a traditional MCMC based approach but it can generate samples from the exact posterior posterior distribution. Fitting such a model using the hybrid approach on a five-dimensional dataset from Simulation 1 with $N = 1000$ takes on average about \underline{\textbf{246 minutes}}. Recall that one iteration of a fully Bayesian MCMC based approach on the same dataset takes on average 45 minutes for one iteration and the number of iterations required are typically in hundreds. When the primary goal is to detect the underlying clusters (which is the case for our the real data analysis), the VGA based approach is sufficient. However,  when the primarily goal is posterior inference, we recommend the hybrid approach as it can better yield an exact posterior similar to the MCMC-EM approach but is computationally efficient. For simulation studies 1 and 2 in which we show parameter recovery, we show parameter estimation using both VGA and the hybrid approach.

\subsubsection{Initialization}
For initialization of $\hat{z}_{ig}$, we used $k$-means clustering \cite{macqueen67,hartigan79} on the estimate of the underlying latent variable $\eta_i$ obtained by first calculating the underlying composition using $\wv_{i}/\sum_{k=1}^{K+1}\wv_{ik}$ for each observation; mapping this composition to the latent variable $\yv_i$ using the additive log-ratio transformation in Equation~\ref{alr}, and transforming the variable to get $\etav_i$ through Equation~\ref{trans}. For initializing the variational parameters for each observation $\wv_i$, we obtain $\etav_i$ first, same as in the $\hat{z}_{ig}$ initialization step. We use this calculated latent variable $\etav_i$ as initialization of $\mv_{ig}$. $V_{ig}$ for each $i$ are initialized as $K+1$ diagonal matrix such that $\tilde{V}_{kk}=1$ for $k=1,\ldots,K$ and $V_{kk}=0$ for $k=K+1$. $\xi_i$'s are initialized using $1$. According to the initialization on the group label $\hat{z}_{ig}$, $\tilde{\muv}_{g}$ and $\tilde{\Sigmav}_{g}$ are initialized as group-specific mean and covariance of $\etav_i$, respectively.

\subsection{Model Selection and Performance Assessment}\label{mixture_assess}
In the clustering context, the number of components $G$ is unknown.  Hence, one typically fits models for a large range of possible $G$ and the number of clusters is then chosen {\it a posteriori} using a model selection criteria. The Bayesian information criterion (BIC)\cite{BIC} is one of the most popular criteria in the model-based clustering literature.\cite{mcnicholas2016}
\begin{equation*}
\text{BIC} \approx -2\tilde{\Lc}+d\log(n),
\end{equation*} where $\tilde{\Lc}$, defined in Equation \ref{vgcomplik}, is the variational Gaussian lower bound of the complete data log likelihood, and $d$ is the number of free parameters in the model. Specifically, when fitting a $G-$ component model, $d=\frac{(K+1)K}{2}\times G+K\times G+G-1$.

When the true class labels are known (e.g., in simulation studies), we assess the performance of our proposed model using the adjusted Rand index (ARI)\cite{ARI}.  It is a measure of the pairwise agreement between the predicted and true classifications such that an $\text{ARI}$ of 1 indicates perfect classification and $0$ indicates that the classification obtained is no better than by chance.\par 
\section{Simulation Studies and Real Data Analysis}\label{results}
To illustrate the performance of our proposed clustering framework, we conducted two sets of simulation studies. For both studies, the $i-$th observed counts data $\Wv_i$ are generated as:
\begin{enumerate}
	\item First, we generate the total counts $\sum_{k=1}^{K+1} W_{ik}$ as a random number from a uniform distribution $U[5000,10000]$.
	\item Given pre-specified group specific parameters $\muv_g$ and $\Sigmav_g$, we transform using Equation~\ref{musigtrans} to get $\tilde \muv_g$ and $\tilde \Sigmav_{g}$ and generate $\etav_i$ from $\normal(\tilde \muv_g,\tilde \Sigmav_g)$.
	\item Based on $\etav_i$, we calculate $\Thetav_i$ using the inverse additive log-ratio transformation $\phi^{-1}$ using Equation~\ref{inv_alr}.
	\item Using $\Thetav_i$ as the underlying composition, together with the total counts generated at the first step, we generate discrete random numbers $\Wv_i$ from multinomial distributions.
	\item To initialize the variational parameters, we need to use the additive log-ratio transformation which takes the log transformation of the observed count for taxa $k$ divided by total count for all taxa for sample $i$. If there are any $0$ in the generated count data, we substitute the $0$ with $1$ for initialization.
\end{enumerate}

We also compared the performance of our proposed model to a Dirichlet mixture model \citep{holmes2012} which is widely used to cluster microbiome data. Implementation of the Dirichlet mixture model is available in the {\sf R} package \texttt{DirichletMultinomial} \citep{Morgan2020}. 

\subsection{Simulation Study 1}
\begin{figure}[t]
\begin{center}
	\includegraphics[width=0.46\textwidth]{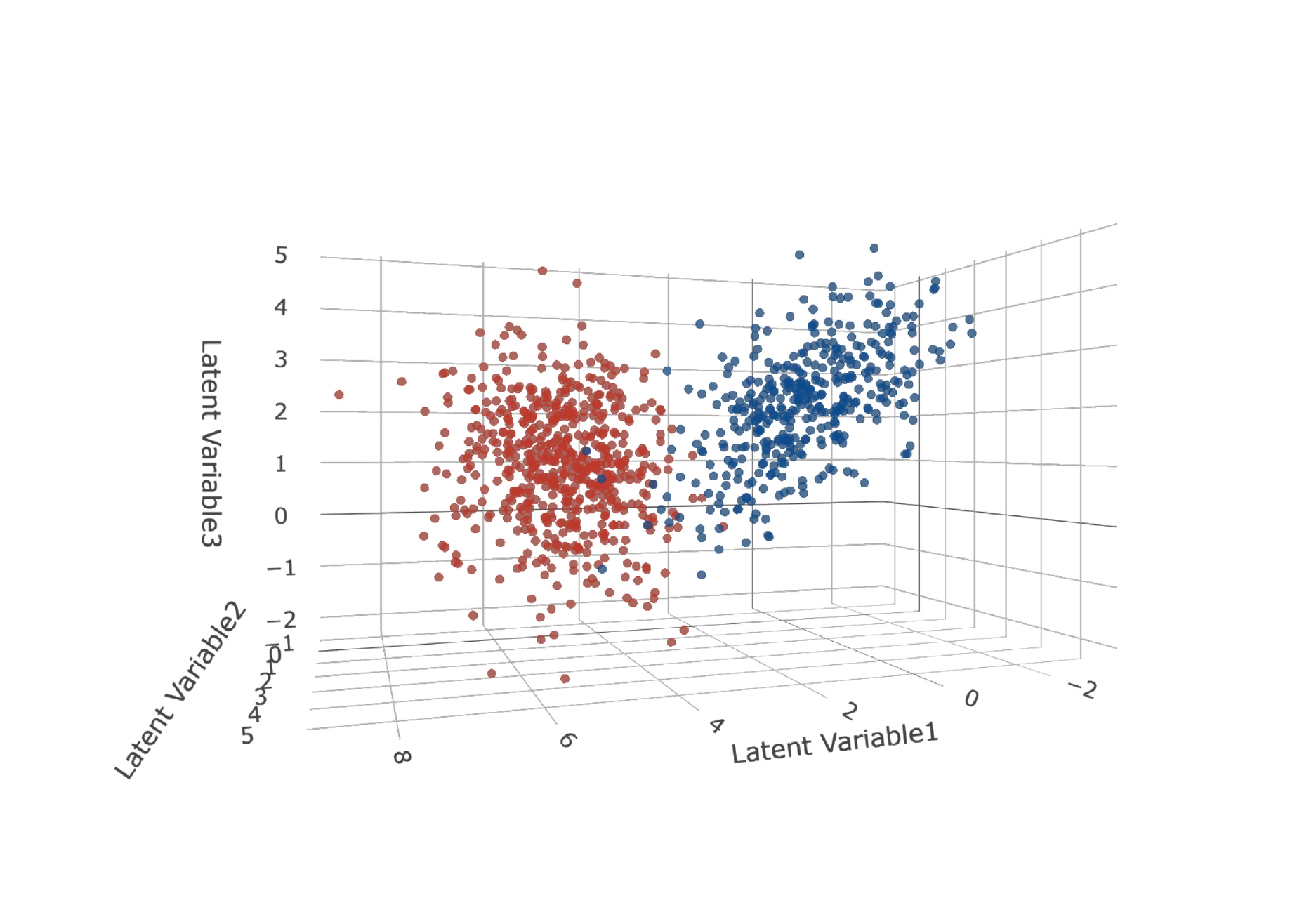}
	\includegraphics[width=0.53\textwidth]{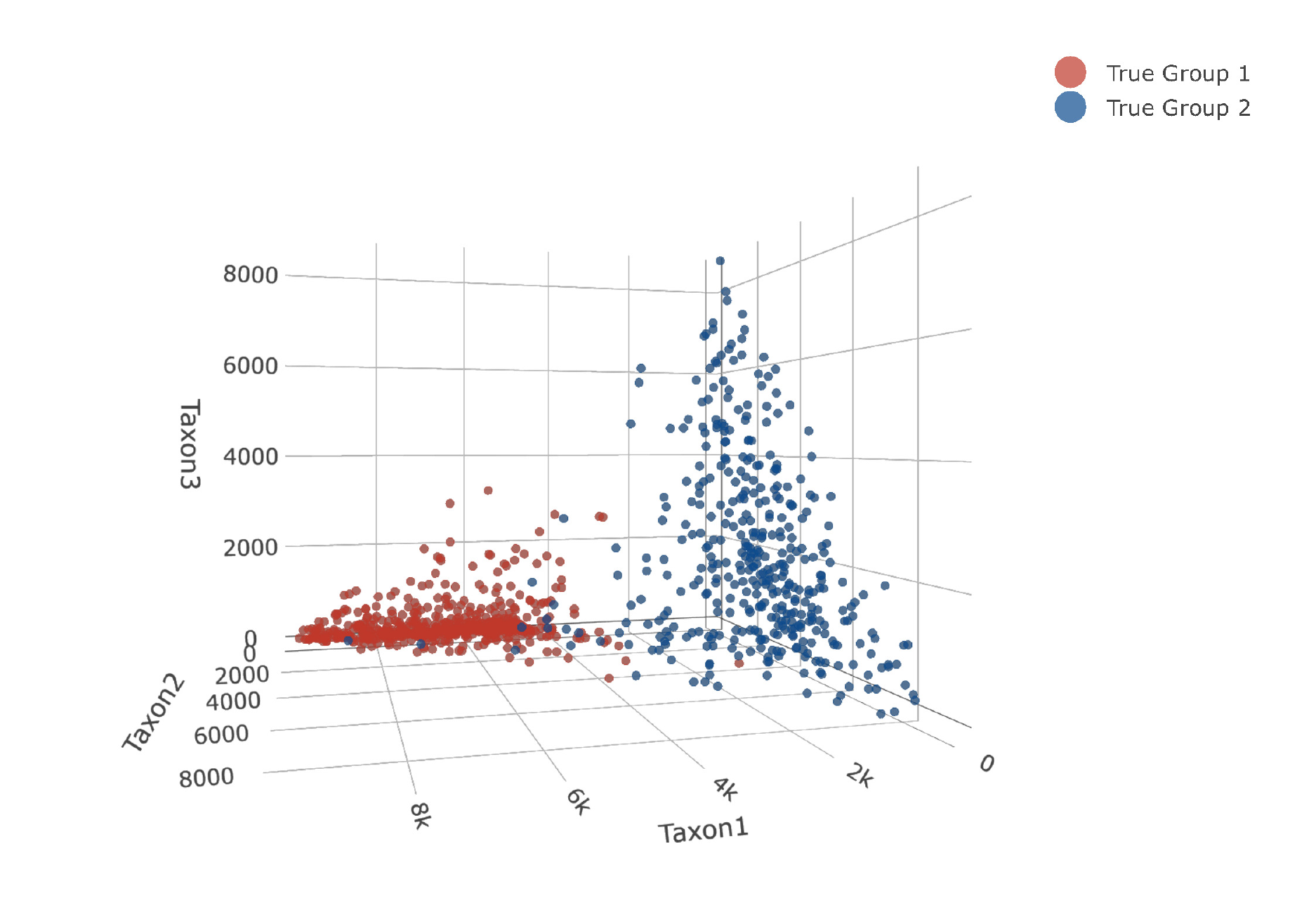}
	\end{center}
	\caption{Three dimensional scatter plot of one of the 100 datasets in Simulation Study 1 highlighting the true labels for the latent variable (left) and the first three columns of the count data (right).\label{fig:sim1}}
\end{figure}
\begin{figure}[t]
	\centerline{\includegraphics[width=342pt,height=9pc]{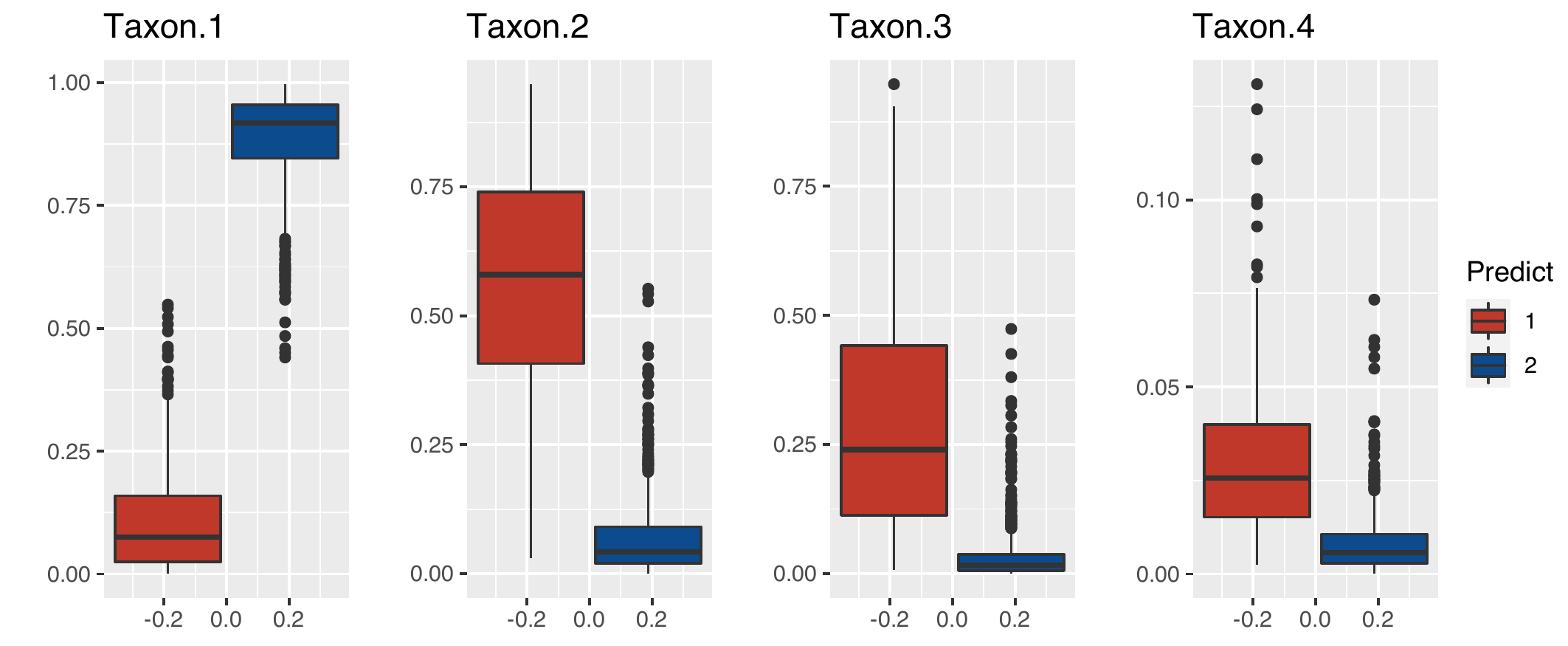}}
	\caption{Boxplots of the relative abundance of observed counts of the four taxa for the predicted clusters for one of the 100 datasets in Simulation ~\ref{fig:sim1}. For this dataset, ARI was 0.95. \label{fig:sim1box}}
\end{figure}

In this simulation study, we generated $100$ datasets where the underlying latent variable $\Yv$ came from two component, three-dimensional multivariate Gaussian distributions with mixing proportions $\piv = (0.6,0.4)$; see Figure~\ref{fig:sim1} (left panel). The first component consists of $n_1=600$ observations and the second component consists of $n_2=400$ observations. The parameters used to generate the datasets are summarized in Table~\ref{tab:s1}. We fitted the models with $G=1,\ldots,5$ on all $100$ datasets. 
In $100$ out of $100$ datasets, BIC selected a two-component model. The models selected by BIC yielded an average ARI $=0.94$ with standard deviation $0.02$. The average and standard deviation of the estimated parameters for the all $100$ datasets using the VGA approach are summarized in Table~\ref{tab:s1} and using the hybrid approach are summarized in Table~\ref{tab:s1_hyb}. Note that the parameter estimation using both approaches are very close to the true value of the parameters. Average computation time for Simulation Study 1 using the proposed VGA approach was 200.97 (sd of 35.39) minutes on a single-core processor. It took additional on average 45.14 (sd of 15.84) minutes for one iteration of the full Bayesian. Thus, the mean computation time using the hybrid approach was 246.11 (sd of 38.57) minutes. Figure~\ref{fig:sim1box} illustrates a clear difference in the distribution of the relative abundance of taxa in the two predicted groups. We also ran the Dirichlet-multinomial mixture model for $G=1:4$ and selected the best model using BIC. In all 100 out of 100 datasets, it selected a $G=4$ model with an average ARI of 0.52 (sd of 0.04). The Dirichlet-multinomial model overestimates the number of components by splitting the true clusters into multiple clusters with some misclassifications among them.

\subsection{Simulation Study 2}
\begin{figure}[t]
\begin{center}
	\includegraphics[width=0.5\textwidth,angle=270]{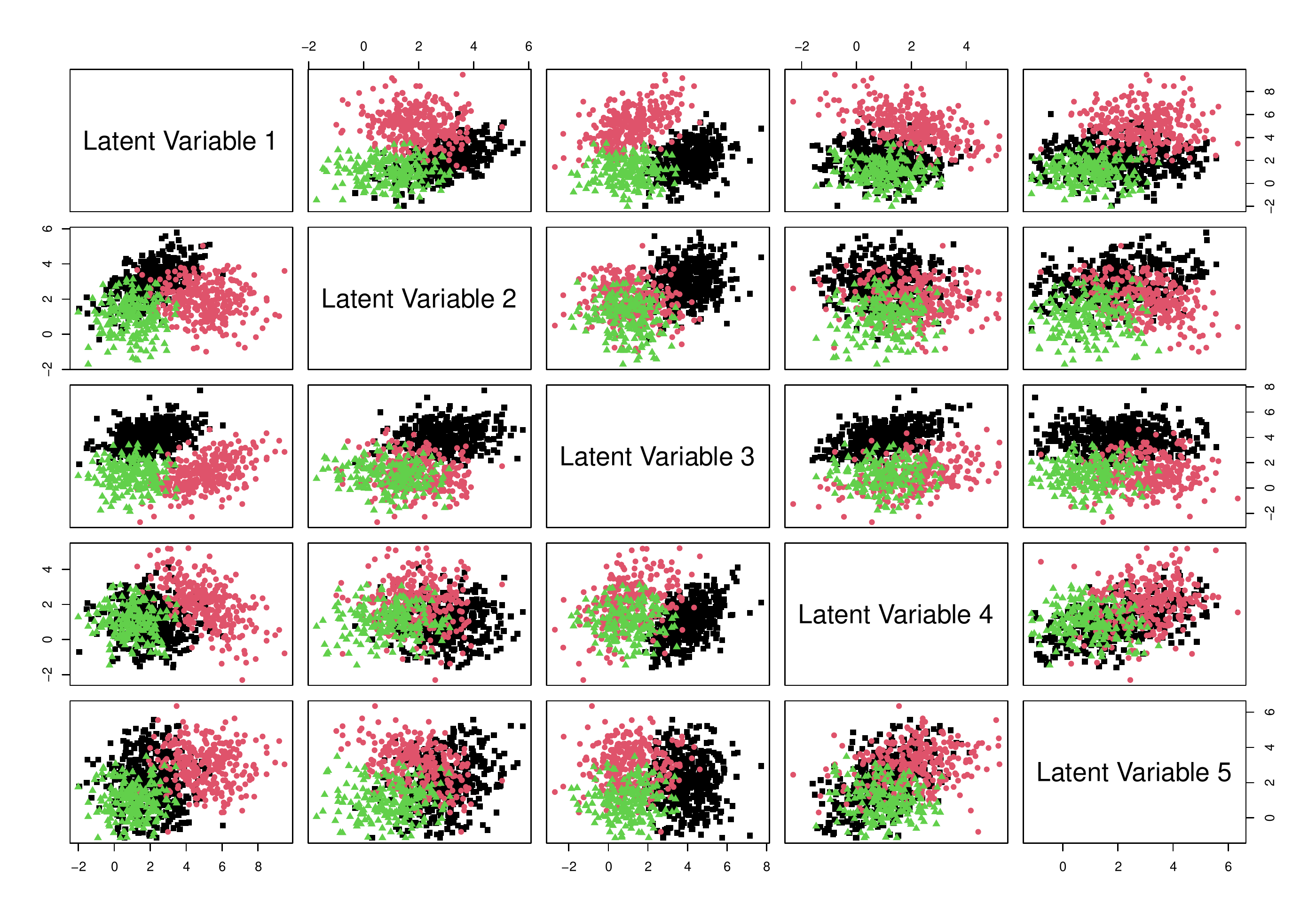}
	\end{center}
	\caption{Pairwise scatter plot of one of the 100 datasets in Simulation Study 2 highlighting the true labels for the latent variables. For this dataset, ARI was 0.95. \label{fig:sim2scatter}}
\end{figure}

In this simulation study, we generated $100$ datasets with the underlying latent variable $\Yv$ from three component five-dimensional multivariate Gaussian distributions (see Figure~\ref{fig:sim2scatter} for the pairwise scatterplot of the underlying latent variable $\Yv_i$).

\begin{figure}[t]
\begin{center}
	\includegraphics[width=0.45\textwidth]{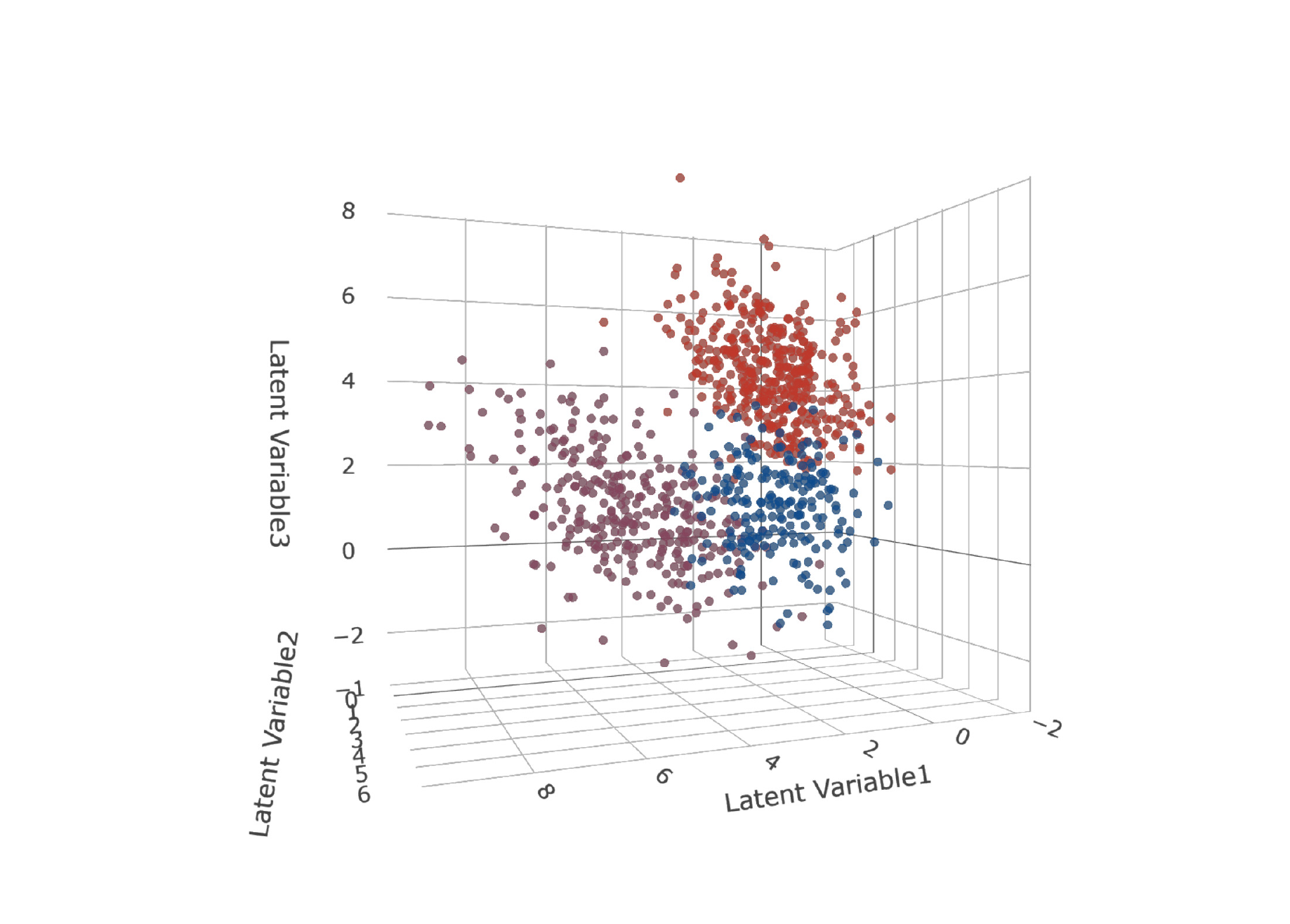}
	\includegraphics[width=0.54\textwidth]{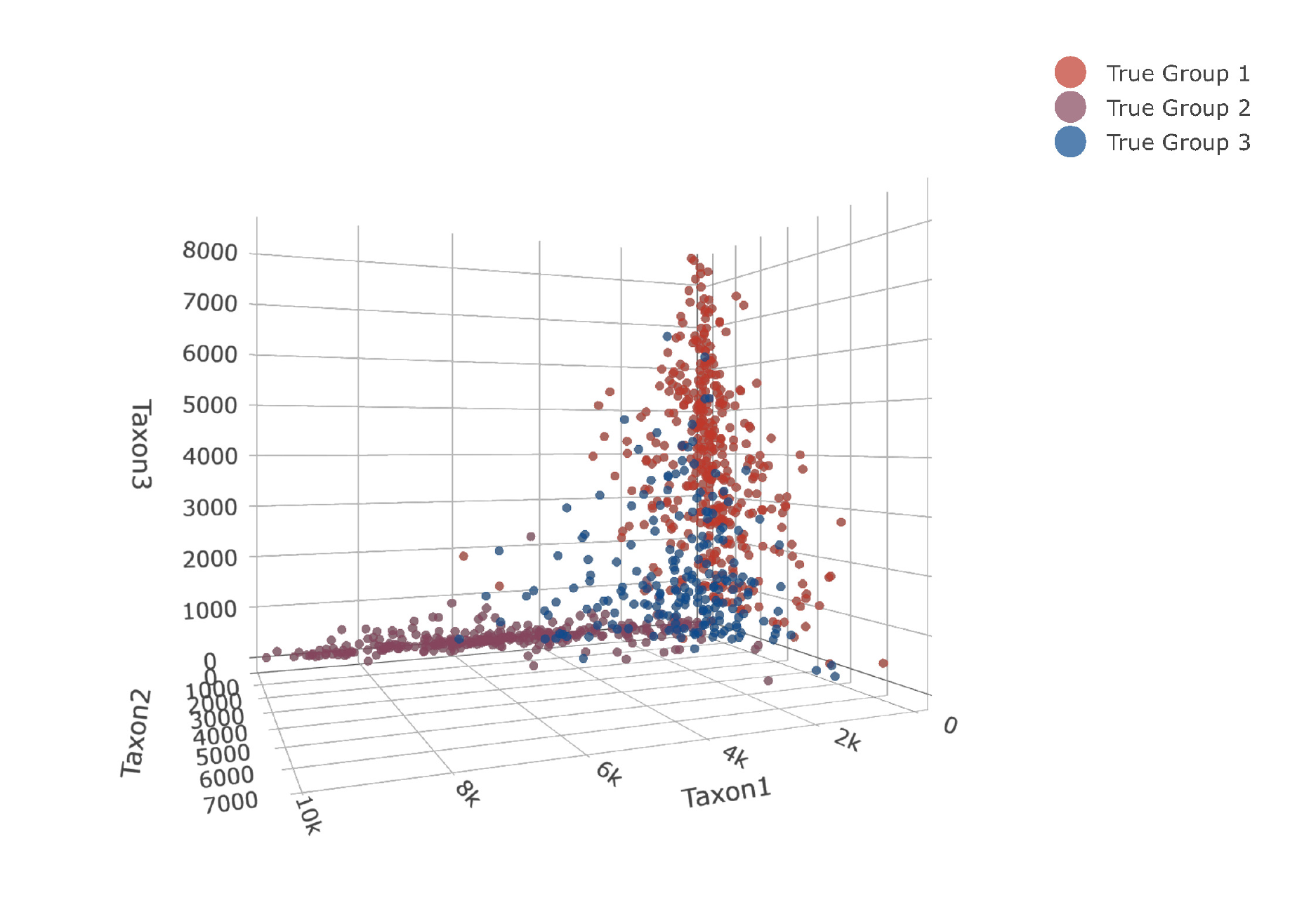}
	\end{center}
	\caption{Three dimensional scatter plot of one of the 100 datasets in Simulation Study 2 highlighting the true labels for the latent variable (left) and the first three columns of the count data (right).\label{fig:sim2}}
\end{figure}

There are $n_1=300$ observations in Group 1, $n_2=400$ observations in Group 1, and $n_3=200$ observations in Group 3. The true parameters are summarized in Table~\ref{tab:s2}. Figure~\ref{fig:sim2} (left panel) shows the first three dimensions of the latent variable $\Yv_i$'s and Figure~\ref{fig:sim2} (right panel) shows the first three dimensions of the observed counts $\Wv_i$'s.  There is a more separation between the groups when visualizing the latent variables as opposed to the observed counts.

\begin{figure}[t]
	\centerline{\includegraphics[width=0.8\textwidth]{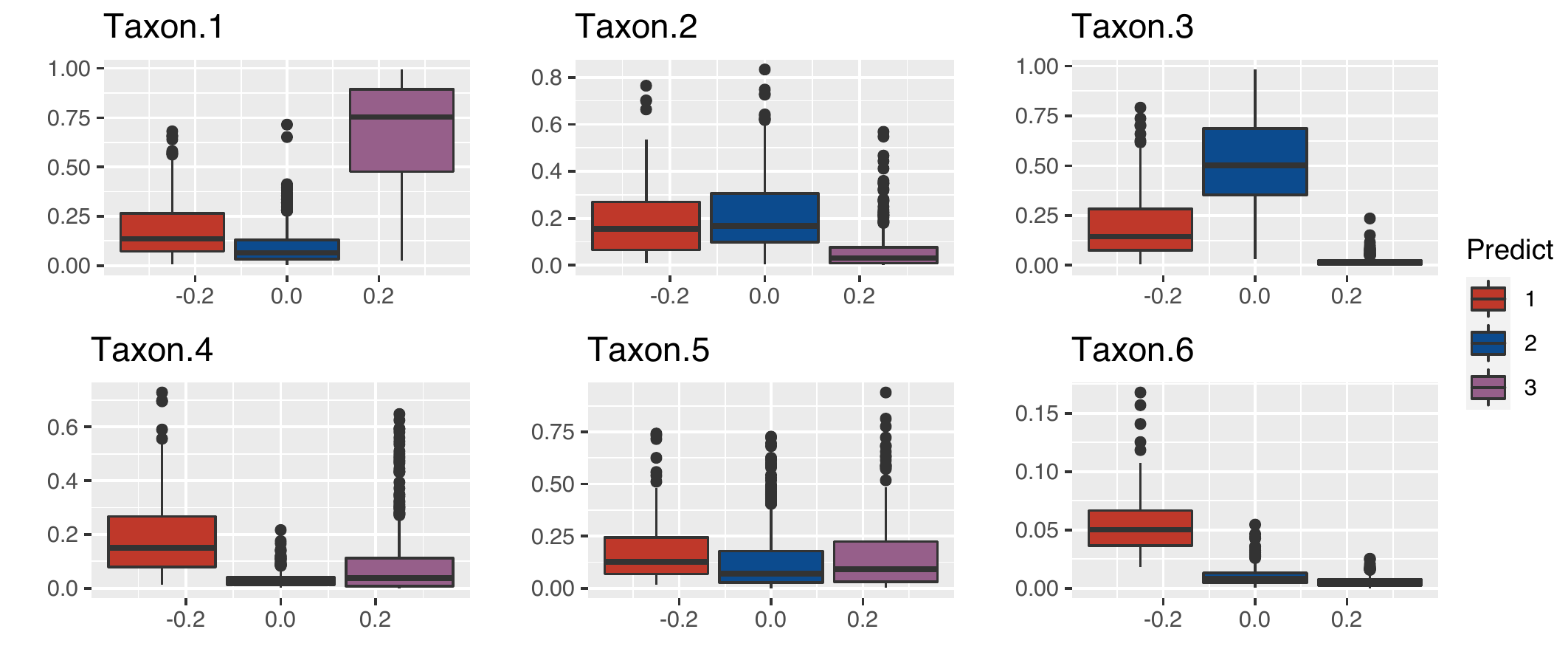}}
	\caption{Boxplots of the relative abundance of observed counts of the four taxa for the predicted clusters for one of the 100 datasets in Simulation ~\ref{fig:sim2}. For this dataset, ARI was 0.95.\label{fig:sim2box}}
\end{figure}

 The proposed algorithm was applied on all $100$ datasets where for each dataset, we fitted the models with for $G=1,\dots,4$. In 94 out of the 100 datasets, a $G=3$ model was selected using the BIC, a $G=2$ was selected for 3 out of the 100 datasets, and $G=4$ models were selected for the remaining 3 datasets. The overall mean ARI for all 100 datasets was 0.93 (sd of 0.06) and the mean ARI for 94 datasets where a three component model was selected was 0.94 (sd of 0.01).  The average and standard deviation of the estimated parameters for the all $100$ datasets using the VGA approach are summarized in Table~\ref{tab:s2} and using the hybrid approach are summarized in Table~\ref{tab:s2_hyb}. Note that the parameter estimation using both approaches are very close to the true value of the parameters. Average computation time for Simulation Study 1 using the proposed VGA approach was 214.91 (sd of 35.91) minutes on a single-core processor. It took additional on average 40.84 (sd of 16.64) minutes for one iteration of the full Bayesian. Thus, the mean computation time for the hybrid approach was 255.74 (sd of 42.70) minutes. Figure~\ref{fig:sim2box} illustrates a clear difference in the distribution of the relative abundance of taxa in the predicted groups. We also ran the Dirichlet-multinomial mixture model for $G=1:4$ and selected the best model using BIC. In all 100 out of 100 datasets, it selected a $G=4$ model with an average ARI of 0.33 (sd of 0.03).

\subsection{Additional Simulation Studies}\label{sec:newsim}
To test the performance of the proposed algorithm on higher dimensional datasets, as well as datasets generated from mixture of Dirichlet-multinomial models, we performed a series of 10 additional simulation studies, each containing 100 datasets, as described below:
\begin{itemize}
	\item Generate 100 datasets from a two-component mixture of logistic normal multinomial models with each of the following:
	\begin{itemize}
		\item $K$, the dimension of the latent variable, being $5,10,$ and $20$;
		\item $n$, the sample size, being $100,200,$ and $500$.
		\item True parameters are the same for different $n$ but same $K$.
	\end{itemize}
	\item Generate 100 datasets from mixture of two-component Dirichlet-multinomial models with dimension 6, and sample size of 200.
\end{itemize}

We ran the proposed algorithm for $G=1:5$ on all datasets and used BIC for model selection. We also applied the Dirichlet-multinomial mixture (DMM) models on these datasets with BIC for model selection. Table~\ref{tab:newsim_select} shows the number of times the correct model ($G=2$) was selected out of 100 datasets for each simulation study, as well as the average ARI with standard deviation, for the proposed algorithm and the Dirichlet-multinomial mixture models. 

When data were generated from a mixture of logistic normal multinomial models, in all simulation scenarios, the proposed algorithm identified the correct number of components for more than 80 times out of 100 with average ARI $\geq 0.96$, except for the case with $K=20, n=100$. Note that in the later case, the number of parameters needed to be estimated is far larger than the sample size. Also, it is observed that, in general, when sample size increases, performance of the proposed algorithm in terms of the number of times it selects the correct model as well as the average ARI also increases. However, the Dirichlet-multinomial mixture model did not perform as well on data simulated from the logistic normal multinomial mixture models. Even in the case of $K = 20, n = 200$, where it correctly selected the two-component model 99 out of 100 times, the average ARI was only 0.6286 with standard deviation of 0.1284. 

When the data was generated from the Dirichlet-multinomial mixture models,  our proposed model is able to recover the underlying cluster in 85 out of the 100 datasets with an average ARI of $0.8957$ and standard deviation of $0.0916$ whereas the Dirichlet-multinomial mixture model was able to recover the underlying cluster structure in all 100 datasets with an average ARI of 0.9589 (sd=0.0456). When performing each simulation study, the computational job was distributed onto a computer cluster, where the proposed algorithm applied on each one of the 100 datasets was run on a one-core slot. Table~\ref{tab:newsim_select} summarizes the average elapsed time for running the proposed algorithm in minutes, with standard deviation. For most cases, it takes the proposed algorithm less than 60 minutes. As the number of observations and the dimensionality of data increases, the time to convergence increased as well.

Table~\ref{tab:newsim_otherG} summarizes how many times each $G=1:5$ were selected by the proposed algorithm. In nine out of the ten studies, our approach was able to identify the correct number of components in at least 85 out of 100 datasets. For the scenario when $K=20$ and $n=100$ (small sample size and high dimensional), both proposed algorithm and the Dirichlet-multinomial mixture model approach favoured a one-component model. We also summarized the average of $L_1$ norm between the true parameters and the estimated values along with the standard errors for the simulations with data generated from mixture of logistic normal multinomial models in Table~\ref{tab:newsim_est}. It shows that, when the dimensionality is low, the proposed algorithm can not only identify the correct underlying group structure but also is able to recover the true parameters well. As the dimensionality increases, the proposed algorithm can still capture the true number of components in the data with high classification accuracy and the estimated central location parameter ($\muv$) is also estimated close to the true value. However, the estimation of the spread parameter ($\Sigmav$) become less precise as dimensionality becomes higher; however, the distance between the true and the estimated parameters decreases as the sample size becomes larger.

\color{black}
\subsection{Real Data Analysis}\label{real_data}
 We applied our proposed algorithms to \textit{four} microbiome datasets from \textit{three} studies:
\begin{itemize}
\item \textbf{The Ferretti 2018 Dataset}:
We applied our algorithm to the Ferretti 2018 dataset \citep{ferretti2018} available through the {\sf R} package \texttt{curatedMetagenomicData} \citep{curated2017} as \texttt{FerrettiP\_2018} dataset.  The study aims to understand the acquisition and development of the infant microbiome and assessed the impact of the maternal microbiomes on the development of infant oral and fecal microbial communities from birth to 4 months of life. Twenty five mother infant pairs who vaginally delivered healthy newborns at full term were recruited for the study. For each mother, stool (proxy for gut microbiome), dorsum tongue swabs (for oral microbiome), vaginal introitus swabs (for vaginal microbiome), intermammary cleft swabs (skin microbiome) and breast milk were obtained. As the DNA extraction from breast milk was not feasible in most cases, they were not analyzed further. All infants were exclusively breastfed  at 3 days, 96\% at 1 month, and 56\% at 4 months and for each newborn, oral cavity and gut samples were taken from birth to up to 4 months. The samples were sequenced using high-resolution shotgun metagenomics \citep{quince2017} with an improved strain-level computational profiling of known and poorly characterized microbiome members \citep{segata2018}. See Ferretti et al (2018) \cite{ferretti2018} for further details. Here, we applied our algorithms to two subsets of the datasets: comparing gut microbiome of the infant with their mothers and comparing oral microbiome of the infants with their mothers.

\noindent \underline{Gut microbiome subset}: This subset of Ferretti 2018 dataset available through {\sf R} package \texttt{curatedMetagenomicData} comprised of  119 samples (23 adults and 96 newborns). As mentioned in  Ferretti et al (2018), stool samples of newborns were taken at five different time points: Day 1, Day 3, Day 7, 1 Month and 4 Months. As repeated measurements at different time points are taken on the same individuals and our model currently cannot model that (violates the independence assumption), we only focus on one time point (i.e., Day 1) for the newborns. Hence, the resulting dataset comprises  of 42 individuals (23 adults and 19 infants). Here, we focused our analysis on the OTU counts at the genus level data.

\noindent \underline{Oral microbiome subset}: This subset of Ferretti 2018 dataset available through {\sf R} package \texttt{curatedMetagenomicData} comprised of 62 samples (23 adults and 39 infants). As mentioned in  Ferreti et al (2018), oral samples of infants were taken at two different time points: Day 1 and Day 3. Here, as the Day 3 had measurements for all 23 newborns, we use the Day 3 measurements for the analysis. The resulting dataset consists of 46 individuals (23 adults and 23 infants). Here, we again focused our analysis on the OTU counts at the genus level data. 

\item \textbf{The Shi 2015 Data}:
We also applied our algorithm to the Shi 2015 dataset \citep{shi2015} available through the {\sf R} package \texttt{curatedMetagenomicData} \citep{curated2017} as \texttt{Shi\_2015} dataset. Periodontitis is a common oral disease that affects about 50\% of the American adults and is associated with alterations in the subgingival microbiome of individual tooth sites.  The study aimed to identify and predict the disease progression using the compositions of the  subgingival microbiome. Samples were collected from 12 healthy individuals with chronic periodontitis from multiples tooth sites per individual, before and after nonsurgical therapy that consisted of scaling and root planing.  Only the samples from the tooth sites that were clinically resolved after the therapy were selected for the study resulting in an average  of two sites per subject. Although samples were obtained from multiple sites of individuals, Shi and colleagues \citep{shi2015} state that individual tooth sites are likely to have independent clinical states and unique microbial communities in subgingival pockets, and therefore, we treated them as independent samples for our analysis. This resulted in a total of 48 samples (24 periodontitis samples and 24 recovered samples). 
\item \textbf{The atlas1006 Data}:
We also applied our algorithm to the atlas1006 dataset available through the {\sf R} package \texttt{microbiome} \citep{lahti2014tipping}. The dataset comprises of microbiome compositions of 1045 western adults with no reported health complications. Covariates information such as age, BMI category, sex, and nationality are also available. The BMI category was underweight ($n=21$), lean ($n$=484), overweight ($n=197$), obese ($n=222$), severe obese ($n=99$) and morbid-obese  ($n=22$). For our analysis, we combined both underweight and lean into one category ``lean" and obese, severe obese, and morbid-obese into one category ``obese", thus resulting in three BMI categories.
\end{itemize}

As our approach is currently not designed for high dimensional data, we first utilized the {\sf R} package \texttt{ALDEx2} \citep{fernandes2013,gloor2015} for differential abundance analysis on the observed genus counts to identify the genera that are different among different groups in the datasets.  This step is analogous to conducting differential expression analysis in RNA-seq studies before performing cluster analysis to identify variables that are group differentiating. As the sample size of \texttt{Ferretti 2018} and \texttt{Shi 2015} datasets were less than 50, we only focused on a small set (top four) of differentially abundant genera and aggregated the remaining genera in a category ``Others" to preserve relative abundance. For the \texttt{atlas1006} dataset which has a sample size of $n=1045$, we used top 20 differentially abundant genera that were differentially abundant in ``obese" and ``lean" category. This ``Other'' genus was then used as the reference level for computing the underlying composition and conducting the additive log-ratio transformation.  We applied our algorithm to this dataset for $G=1$ to $4$ on all datasets and used BIC for model selection. We also ran the Dirichlet-multinomial mixture model with the same set of genera for $G=1$ to $4$ on all datasets and utilized BIC for model selection. Summary of the clustering performances are provided in Table \ref{ctRD1}. In two out of the four datasets (i.e., the \texttt{Ferretti} gut microbiome dataset and the \texttt{Shi} dataset) our approach outperforms the Dirichlet-multinomial mixture models while both approaches provides a perfect classification on  the \texttt{Ferretti} oral microbiome dataset. It is important to note that for the \texttt{atlas1006} datasets, both approaches fail to recover the underlying group structure. 

To visualize the true and recovered cluster structure, we conducted principal component analysis using the transformed variable $\mathbf{Y}$. 

\begin{figure}
\begin{center}
\includegraphics[angle=270,width=\textwidth]{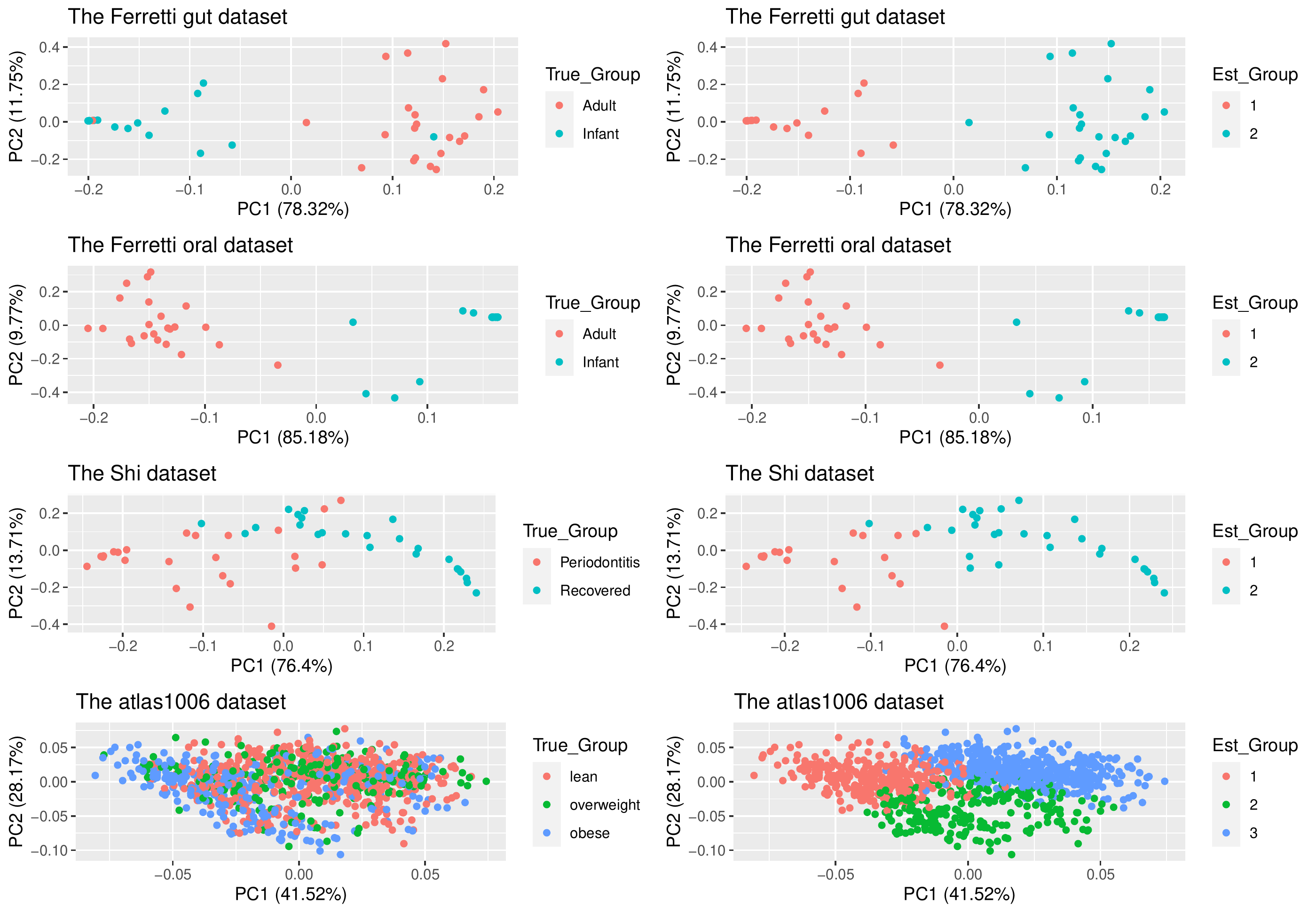}
	\caption{Visualization of the true and estimated clusters on all four real datasets using principal component analysis.}\label{pca}
\end{center}
\end{figure}

Two observations were misclassified in the \texttt{Ferretti} gut dataset using our proposed algorithm, and as can be seen in Figure \ref{pca}, the misclassified infant had a similar microbiome composition to the adult and the misclassified adult had a similar microbiome composition to the infant. Similarly, as seen in Figure \ref{pca}, most of the misclassified observations for the \texttt{Shi} dataset also had a similar microbiome composition to the group they were assigned. Shi and colleagues \citep{shi2015} also utilized the microbiome profiles  to classify the clinical state. They performed a supervised classification using the weighted gene-voting algorithm and leave-one-out cross-validation yielding 33 out of 48 samples as correctly classified into the respective clinical state (correct classification rate of 68.75\%) whereas 6 samples were misclassified into the incorrect clinical state and 9 samples were not assigned to any clinical state due to low prediction strength. Note that in supervised classification, the group labels are used to build a predictive model which is then used to make predictions on new or ``future" observations. Here, we achieved a correction classification rate of 85.42\% (i.e., 41/48 correct classification). On the \texttt{atlas1006} dataset, recall that both the proposed algorithm and the Dirichlet-multinomial approach did not perform well in recovering the underlying groups based on the BMI. Visualization of the dataset using the principal components show that there is in fact not a clear separation in the microbiome compositions between the various BMI categories in the dataset. Although the recovered cluster structure does not correspond to the known BMI categories, Figure \ref{pca} shows that the proposed approach is able to recover homogenous clusters. We do not have further information to investigate what the recovered cluster structure can yield insight into

Based on suggestion from a reviewer, an alternate approach to using the differentially abundant taxa by using the most abundant taxa instead was investigated. To keep the dimensionality the same, for \texttt{Ferretti} and \texttt{Shi} datasets, we used the top 4 most abundant taxa and for the \texttt{atlas1006} dataset, we used the top 20 most abundant taxa. We ran our proposed approach for $G=1,\ldots,4$ and used BIC for model selection. Our approach selected a one-component model (ARI of 0) for \texttt{Ferretti} gut microbiome dataset, a two-component model (ARI of 0.36, 9/46 misclassification) for \texttt{Ferretti} oral microbiome dataset, a two-component model (ARI of 0.05,18/48 misclassification) for the \texttt{Shi} dataset, and a two-component model (ARI of 0.002) for \texttt{atlas1006} dataset. Note that using the most abundant genera for all four dataset resulted in a far lower ARI compared to using the most differentially abundant genera.

\section{Conclusion}\label{discussion}
A model-based clustering framework for microbiome compositional data is developed using a mixture of logistic normal multinomial models. The novelty of this work is multi-fold. Previous work \cite{xia2013} has indicated that the logistic normal multinomial models can model the dependency of the bacterial composition in a microbiome compositional data in a more flexible way than the commonly used Dirichlet-multinomial models. The latent variables in the logistic normal multinomial model are assumed to follow a multivariate Gaussian distribution and a closed form expression of the log-likelihood or posterior distributions of the latent variables do not exist. Hence, prior work on model fitting relied on Markov chain Monte Carlo (MCMC) sampling techniques that come with heavy computational burden. This is compounded in the clustering context where MCMC sampling needs to be utilized at every iteration of the variant of EM algorithm that is typically utilized for parameter estimation. Here, we employed a variational Gaussian approximation to the posterior distribution of the latent variable and implemented a generalized EM algorithm that does not rely on MCMC sampling thus making it feasible to extend these models for clustering. This also opening up the possibility of efficiently scaling and extending these models a high dimensional setting.\par
Through simulation studies, we have shown that the proposed algorithm delivers accurate parameter recovery and good clustering performance. The proposed method is also illustrated on four real datasets in Section~\ref{real_data} where we demonstrate that the proposed models can recover the underlying cluster (group) structure in the real data.
While in the datasets with small sample size, we focus on small dimensional data by data aggregation to most differentially abundant genera in real data analysis, for larger dataset, more taxa can be used. Because of adopting an underlying Gaussian distribution, the number of parameters in the covariance matrix alone grows quadratically with $K$. Thus, in high dimensional datasets with small sample size, estimating $\boldsymbol{\Sigma}^{-1}$ becomes more challenging as it can lead to degenerate solutions and a host of other issues related to model convergence and fitting while using a traditional maximum likelihood based expectation-maximization approach. This a well-known issue with Gaussian mixture models and are typically dealt with either variable/feature selection or dimension reduction. Feature selection typically eliminates the redundant or irrelevant variables and reduce computational cost, provide a better understanding of data and improve predictions \citep{haq2019}. ALDEx2 utilized here is a widely used variable/feature selection technique specifically designed for microbiome data that identifies taxa that are differentially abundance in different conditions. Through a comparative study of ALDEx2 with other approaches commonly used for differential abundance analysis, Quinn et al. \cite{quinn2018} showed that ALDEx2 has high precision (i.e., few false positives) across different scenarios. However, information on the group structure or conditions may not be available a-priori. In such case, one may conduct feature selection by selecting the top few most abundant taxa and collapsing low-abundant taxa into one category ``Others" to preserve the compositional nature of the data. Alternately, mixtures of logistic multinomial models can be extended to high-dimensional data by introducing subspace clustering techniques through the latent variable \cite{mcnicholas2008, mcnicholas2010, bouveyron2014s}. This will be the topic of some future work. Additionally, it has been well-established that different environmental or biological covariates can affect the microbiome compositions. Some future work will also focus on developing a mixture of logistic normal multinomial regression models to investigate the relationship of biological/environmental covariates with the microbiome compositions within each clusters. 

\noindent \textbf{Data Availability Statement}: The datasets used in this manuscript are publicly available from the {\sf R} package \texttt{curatedMetagenomicData} and \texttt{microbiome}. 

\noindent \textbf{Funding Acknowledgement}:
This work was supported by Collaboration Grants for Mathematicians by Simons Foundation, NSERC Discovery Grant, and Canada Research Chair Program. (Subedi).


\begin{center}
	\begin{table}[!ht]%
		\centering
		\caption{True and estimated parameters along with the standard deviations from the one hundred datasets for the latent variable parameters in Simulation Study 1 using VGA approach; average ARI$=0.94$ $(0.02)$.\label{tab:s1}}%
		\begin{tabular*}{500pt}{@{\extracolsep\fill}lcc@{\extracolsep\fill}}
			\toprule
			&\multicolumn{2}{@{}c@{}}{\textbf{Component 1 ($n=600$)}}\\
			\cmidrule{2-3}
			\textbf{Parameter} & \textbf{True}  & \textbf{Average of the estimates (sd)}\\
			\midrule
			$\muv$ &  $[5,\quad 2,\quad 1]$  & $[5.00 (0.05),\quad 2.00 (0.05),\quad 1.00 (0.04)]$\\[5pt]
			$\Sigmav$ &$\begin{bmatrix} 1&0.4&0\\0.4&1.2&-0.5\\0&-0.5&1\end{bmatrix}$ & $\begin{bmatrix}
			1.01(0.07)&0.42(0.06)&-0.01(0.05)\\0.42(0.06)&1.21(0.08)&-0.50(0.06)\\-0.01(0.05)&-0.50(0.06)&0.98(0.07)\end{bmatrix}$\\
			\midrule
			&\multicolumn{2}{@{}c@{}}{\textbf{Component 2 ($n=400$)}} \\
			\cmidrule{2-3}
			\textbf{Parameter} & \textbf{True}  & \textbf{Average of the estimates (sd)}\\
			\midrule
			$\muv$ & $[1,\quad 3,\quad 2]$ & $[1.01 (0.07),\quad 3.00 (0.05),\quad 2.00 (0.05)]$\\[5pt]
			$\Sigmav$ &$\begin{bmatrix}1.4&0.2&-0.65\\0.2&1&0\\-0.65&0&1\end{bmatrix}$ & $\begin{bmatrix}
			1.41(0.12)&0.20(0.07)&-0.65(0.08)\\0.20(0.07)&1.00(0.08)&-0.01(0.05)\\-0.65(0.08)&-0.01(0.05)&0.97(0.08)\end{bmatrix}$\\
			\bottomrule
		\end{tabular*}
	\end{table}
\end{center}

\begin{center}
	\begin{table}[!ht]%
		\centering
		\caption{True and estimated parameters along with the standard deviations from the one hundred datasets for the latent variable parameters in Simulation Study 1 using the hybrid approach.\label{tab:s1_hyb}}%
		\begin{tabular*}{500pt}{@{\extracolsep\fill}lcc@{\extracolsep\fill}}
			\toprule
			&\multicolumn{2}{@{}c@{}}{\textbf{Component 1 ($n=600$)}}\\
			\cmidrule{2-3}
			\textbf{Parameter} & \textbf{True}  & \textbf{Average of the estimates (sd)}\\
			\midrule
			$\muv$ &  $[5,\quad 2,\quad 1]$  & $[4.99 (0.05),\quad 2.00 (0.05),\quad 0.99 (0.04)]$\\[5pt]
			$\Sigmav$ &$\begin{bmatrix} 1&0.4&0\\0.4&1.2&-0.5\\0&-0.5&1\end{bmatrix}$ & $\begin{bmatrix}
			0.98 (0.07)&0.41 (0.05)&0.00 (0.05)\\0.41 (0.05)&1.21 (0.08)&-0.50 (0.06)\\0.00 (0.05)&-0.50 (0.06)&0.99 (0.06)\end{bmatrix}$\\
			\midrule
			&\multicolumn{2}{@{}c@{}}{\textbf{Component 2 ($n=400$)}} \\
			\cmidrule{2-3}
			\textbf{Parameter} & \textbf{True}  & \textbf{Average of the estimates (sd)}\\
			\midrule
			$\muv$ & $[1,\quad 3,\quad 2]$ & $[0.98 (0.07),\quad 3.01 (0.05),\quad 2.02 (0.05)]$\\[5pt]
			$\Sigmav$ &$\begin{bmatrix}1.4&0.2&-0.65\\0.2&1&0\\-0.65&0&1\end{bmatrix}$ & $\begin{bmatrix}
			1.32 (0.11)&0.23 (0.07)&-0.59 (0.07)\\0.23 (0.07)&1.00 (0.08)&-0.01 (0.05)\\-0.59 (0.07)&-0.01 (0.05)&0.95 (0.08)\end{bmatrix}$\\
			\bottomrule
		\end{tabular*}
	\end{table}
\end{center}

\begin{center}
	\begin{table}[!ht]%
		\centering
		\caption{True and Estimated Values for the Latent Variable Parameters in Simulation Study 2 using VGA approach; Average ARI$=0.94 (0.01)$.\label{tab:s2}}%
		\scalebox{0.8}{\begin{tabular*}{610pt}{@{\extracolsep\fill}lcc@{\extracolsep\fill}}
			\toprule
			&\multicolumn{2}{@{}c@{}}{\textbf{Component 1 ($n=300$)}}\\
			\cmidrule{2-3}
			\textbf{Parameter} & \textbf{True}  & \textbf{Estimated (sd)}\\
			\midrule
			$\muv$ &  $[5,\quad 2,\quad 1,\quad 2,\quad 3]$  & $[5.01 (0.09),\quad 2.01 (0.06),\quad 1.00 (0.08),\quad 2.00 (0.07),\quad 3.01 (0.06)]$\\[5pt]
			$\Sigmav$ &$\begin{bmatrix}2&-0.2&0.8&-1&0\\-0.2&1&-0.2&0&-0.4\\0.8&-0.2&1.4&0.6&0\\-1&0&0.6&1.6&0.2\\0&-0.4&0&0.2&1.2\end{bmatrix}$ & $\begin{bmatrix}
			2.03(0.17)&-0.17(0.08)&0.82(0.13)&-0.99(0.11)&0.02(0.10)\\-0.17(0.08)&0.98(0.08)&-0.18(0.07)&-0.01(0.08)&-0.39(0.07)\\
0.82(0.13)&-0.18(0.07)&1.42(0.13)&0.61(0.10)&0.00(0.08)\\
-0.99(0.11)&-0.01(0.08)&0.61(0.10)&1.61(0.14)&0.19(0.07)\\0.02(0.10)&-0.39(0.07)&0.00(0.08)&0.19(0.07)&1.22(0.10)\end{bmatrix}$ \\
			\midrule
			&\multicolumn{2}{@{}c@{}}{\textbf{Component 2 ($n=400$)}} \\
			\cmidrule{2-3}
			\textbf{Parameter} & \textbf{True}  & \textbf{Estimated (sd)}\\
			\midrule
			$\muv$ &$[2,\quad 3,\quad 4,\quad 1,\quad 2]$&$[1.99 (0.07),\quad 2.99 (0.06),\quad 3.99 (0.06),\quad 1.00 (0.06),\quad 2.00 (0.07)]$\\[5pt]
			$\Sigmav$ &$\begin{bmatrix}1.4&0.65&0.4&0&0\\0.65&1&0.2&0&0.4\\0.4&0.2&1&0.6&0\\0&0&0.6&1.2&0.8\\0&0.4&0&0.8&2\end{bmatrix}$ & $\begin{bmatrix}
			1.36(0.11)&0.62(0.07)&0.38(0.09)&-0.01(0.08)&-0.01(0.09)\\
			0.62(0.07)&0.98(0.09)&0.19(0.07)&-0.01(0.06)&0.39(0.08)\\
0.38(0.09)&0.19(0.07)&0.99(0.09)&0.60(0.07)&-0.01(0.08)\\
-0.01(0.08)&-0.01(0.06)&0.60(0.07)&1.19(0.08)&0.78(0.09)\\
-0.01(0.09)&0.39(0.08)&-0.01(0.08)&0.78(0.09)&1.97(0.15)\end{bmatrix}$ \\
			\midrule
			&\multicolumn{2}{@{}c@{}}{\textbf{Component 3 ($n=200$)}} \\
			\cmidrule{2-3}
			\textbf{Parameter} & \textbf{True}  & \textbf{Estimated (sd)}\\
			\midrule
			$\muv$ &$[1,\quad 1,\quad 1,\quad 1, \quad 1]$&$[1.00(0.08),\quad 0.98(0.07),\quad 1.00(0.09),\quad 1.01(0.09),\quad 1.01(0.08)]$\\[5pt] 
		    $\Sigmav$ &$\begin{bmatrix}1&0&0&0&0\\0&1&0&0&0\\0&0&1&0&0\\0&0&0&1&0\\0&0&0&0&1\end{bmatrix}$ & $\begin{bmatrix}
			1.00(0.13)&-0.01(0.08)&-0.02(0.08)&-0.02(0.08)&-0.02(0.07)\\
			-0.01(0.08)&0.97(0.11)&-0.01(0.09)&0.00(0.08)&0.00(0.07)\\
			-0.02(0.08)&-0.01(0.09)&0.99(0.12)&-0.01(0.08)&0.00(0.08)\\
			-0.02(0.08)&0.00(0.08)&-0.01(0.08)&0.98(0.11)&0.00(0.07)\\
			-0.02(0.07)&0.00(0.07)&0.00(0.08)&0.00(0.07)&0.98(0.10)\end{bmatrix}$\\
			\bottomrule
		\end{tabular*}
		}
	\end{table}
\end{center}

\begin{center}
	\begin{table}[!ht]%
		\centering
		\caption{True and estimated parameters (mean (sd)) for Simulation Study 2 using the hybrid approach.\label{tab:s2_hyb}}%
		\scalebox{0.8}{\begin{tabular*}{610pt}{@{\extracolsep\fill}lcc@{\extracolsep\fill}}
			\toprule
			&\multicolumn{2}{@{}c@{}}{\textbf{Component 1 ($n=300$)}}\\
			\cmidrule{2-3}
			\textbf{Parameter} & \textbf{True}  & \textbf{Estimated (sd)}\\
			\midrule
			$\muv$ &  $[5,\quad 2,\quad 1,\quad 2,\quad 3]$  & $[5.00 (0.09),\quad 2.00 (0.06),\quad 0.99 (0.08),\quad 2.00 (0.08),\quad 3.00 (0.06)]$\\[5pt]
			$\Sigmav$ &$\begin{bmatrix}2&-0.2&0.8&-1&0\\-0.2&1&-0.2&0&-0.4\\0.8&-0.2&1.4&0.6&0\\-1&0&0.6&1.6&0.2\\0&-0.4&0&0.2&1.2\end{bmatrix}$ & $\begin{bmatrix}
			2.01 (0.17)&-0.18 (0.08)&0.80 (0.13)&-0.99 (0.11)&0.01 (0.10)\\-0.18 (0.08)&0.98 (0.08)&-0.19 (0.07)&0.00 (0.08)&-0.39 (0.07)\\
0.80 (0.13)&-0.19 (0.07)&1.40 (0.12)&0.61 (0.09)&0.00 (0.08)\\
-0.99 (0.11)&0.00 (0.08)&0.61 (0.09)&1.61 (0.13)&0.20 (0.07)\\0.01 (0.10)&-0.39 (0.07)&0.00 (0.08)&0.20 (0.07)&1.22 (0.10)\end{bmatrix}$ \\
			\midrule
			&\multicolumn{2}{@{}c@{}}{\textbf{Component 2 ($n=400$)}} \\
			\cmidrule{2-3}
			\textbf{Parameter} & \textbf{True}  & \textbf{Estimated (sd)}\\
			\midrule
			$\muv$ &$[2,\quad 3,\quad 4,\quad 1,\quad 2]$&$[1.99 (0.07),\quad 2.99 (0.06),\quad 3.99 (0.06),\quad 1.00 (0.06),\quad 2.00 (0.07)]$\\[5pt]
			$\Sigmav$ &$\begin{bmatrix}1.4&0.65&0.4&0&0\\0.65&1&0.2&0&0.4\\0.4&0.2&1&0.6&0\\0&0&0.6&1.2&0.8\\0&0.4&0&0.8&2\end{bmatrix}$ & $\begin{bmatrix}
			1.36(0.11)&0.62(0.07)&0.38(0.09)&0.00(0.08)&-0.01(0.09)\\
			0.62(0.07)&0.98(0.08)&0.18(0.07)&-0.01(0.06)&0.39(0.08)\\
0.38(0.09)&0.18(0.07)&0.98(0.09)&0.60(0.07)&-0.01(0.08)\\
0.00(0.08)&-0.01(0.06)&0.60(0.07)&1.20(0.08)&0.79(0.09)\\
-0.01(0.09)&0.39(0.08)&-0.01(0.08)&0.79(0.09)&1.98(0.15)\end{bmatrix}$ \\
			\midrule
			&\multicolumn{2}{@{}c@{}}{\textbf{Component 3 ($n=200$)}} \\
			\cmidrule{2-3}
			\textbf{Parameter} & \textbf{True}  & \textbf{Estimated (sd)}\\
			\midrule
			$\muv$ &$[1,\quad 1,\quad 1,\quad 1, \quad 1]$&$[0.99(0.08),\quad 0.96(0.08),\quad 0.97(0.09),\quad 1.01(0.09),\quad 1.01(0.09)]$\\[5pt] 
		    $\Sigmav$ &$\begin{bmatrix}1&0&0&0&0\\0&1&0&0&0\\0&0&1&0&0\\0&0&0&1&0\\0&0&0&0&1\end{bmatrix}$ & $\begin{bmatrix}
			0.99(0.13)&-0.01(0.08)&-0.01(0.08)&-0.02(0.08)&-0.02(0.07)\\
			-0.01(0.08)&0.96(0.11)&-0.04(0.09)&0.02(0.08)&-0.01(0.07)\\
			-0.01(0.08)&-0.04(0.09)&0.91(0.11)&0.02(0.07)&0.00(0.07)\\
			-0.02(0.08)&0.02(0.08)&0.02(0.07)&0.98(0.11)&0.00(0.07)\\
			-0.02(0.07)&-0.01(0.07)&0.00(0.07)&0.00(0.07)&0.97(0.10)\end{bmatrix}$\\
			\bottomrule
		\end{tabular*}
		}
	\end{table}
\end{center}

\begin{center}
	\begin{table}[!ht]%
		\centering
		\caption{Summary of the number of times the correct model is selected along with the average ARI (with standard deviation) and average time per simulation (in minutes; with standard deviation) for completion for the 100 datasets for the 10 simulation studies described in Section~\ref{sec:newsim}.\label{tab:newsim_select}}%
		\scalebox{0.8}{\begin{tabular*}{610pt}{@{\extracolsep\fill}lrcrcrr@{\extracolsep\fill}}
				\toprule
				&\multicolumn{3}{@{}c@{}}{\textbf{Proposed algorithm}}&\multicolumn{2}{@{}c@{}}{\textbf{Dirichlet-multinomial mixture model}}\\
				\cmidrule{2-4}\cmidrule{5-6}
				\textbf{Simulation setting}& \textbf{Average time (sd)} & \textbf{Correct G}  & \textbf{ARI (sd)}& \textbf{Correct G}  & \textbf{ARI (sd)}\\
				\midrule
				K=5, n=100	&	10.8910 (1.2946)	&	92	&	0.9815 (0.0310)	&	1	&	0.1101 (0.0000)	\\
				\midrule											
				K=5, n=200	&	21.8986 (2.6654)	&	98	&	0.9897 (0.0152)	&	5	&	0.0234 (0.0266)	\\
				\midrule											
				K=5, n=500	&	36.3913 (24.4512)	&	99	&	0.9898 (0.0089)	&	3	&	0.0086 (0.0087)	\\
				\midrule											
				K=10, n=100	&	13.4434 (1.3011)	&	95	&	0.9578 (0.0499)	&	64	&	0.5753 (0.1159)	\\
				\midrule											
				K=10, n=200	&	27.3833 (2.7137)	&	100	&	0.9849 (0.0274)	&	88	&	0.5985 (0.0728)	\\
				\midrule											
				K=10, n=500	&	80.8086 (11.4708)	&	100	&	0.9962 (0.0080)	&	0	&	N/A	\\
				\midrule											
				K=20, n=100	&	20.9326 (1.8701)	&	23	&	0.7771 (0.0788)	&	20	&	0.7037 (0.1152)	\\
				\midrule											
				K=20, n=200	&	34.2703 (19.5014)	&	99	&	0.9942 (0.0189)	&	99	&	0.6286 (0.1284)	\\
				\midrule											
				K=20, n=500	&	140.9436 (49.8864)	&	100	&	0.9996 (0.0021)	&	0	&	N/A	\\
				\midrule											
				DMM (k=5, n=200)	&	28.8031 (10.6174)	&	85	&	0.8957 (0.0916)	&	100	&	0.9589 (0.0456)	\\
				\bottomrule
		\end{tabular*}}
	\end{table}
\end{center}

\begin{center}
	\begin{table}[!ht]%
		\centering
		\caption{ Summary of the number of times various $G$ are selected for simulation studies described in Section~\ref{sec:newsim}.\label{tab:newsim_otherG}}%
		\scalebox{0.85}{\begin{tabular*}{575pt}{@{\extracolsep\fill}lccccccccc@{\extracolsep\fill}}
				\toprule
				&\multicolumn{4}{@{}c@{}}{\textbf{ Proposed algorithm}}&\multicolumn{5}{@{}c@{}}{\textbf{Dirichlet-multinomial mixture model}}\\
				\cmidrule{2-5} \cmidrule{6-10} 
				\textbf{Simulation setting} & \textbf{$G=1$}  & \textbf{$G=2$}& \textbf{$G=3$}  & \textbf{$G=4$}& \textbf{$G=1$}  & \textbf{$G=2$}& \textbf{$G=3$}  & \textbf{$G=4$}& \textbf{$G=5$}\\
				\midrule
				K=5, n=100	&	1	&	92	&	7	&		&	99	&	1	&		&		&		\\
				K=5, n=200	&		&	98	&	2	&		&	94	&	5	&		&	1	&		\\
				K=5, n=500	&		&	99	&	1	&		&	1	&	3	&	40	&	26	&	30	\\
				\midrule																			
				K=10, n=100	&		&	98	&	2	&		&	26	&	64	&	10	&		&		\\
				K=10, n=200	&		&	100	&		&		&		&	88	&	10	&	1	&	1	\\
				K=10, n=500	&		&	100	&		&		&		&	0	&		&	17	&	83	\\
				\midrule																			
				K=20, n=100	&	77	&	23	&		&		&	80	&	20	&		&		&		\\
				K=20, n=200	&	1	&	99	&		&		&		&	99	&		&		&		\\
				K=20, n=500	&		&	100	&		&		&		&	0	&	30	&	34	&	36	\\
				\midrule																			
				DMM (k=5, n=200)	&		&	85	&	14	&	1	&		&	100	&		&		&\\
				\bottomrule
		\end{tabular*}}
	\end{table}
\end{center}

\begin{center}
	\begin{table}[!ht]%
		\centering
		\caption{Average $L_1$ norm with standard error of the true parameters and estimated values for simulation studies described in Section~\ref{sec:newsim}.\label{tab:newsim_est}}%
		\scalebox{0.85}{\begin{tabular*}{575pt}{
		@{\extracolsep\fill}lccc@{\extracolsep\fill}}
				\toprule
				&\multicolumn{3}{@{}c@{}}{\textbf{Component 1 ($\pi_1 = 0.5$)}}\\
				\cmidrule{2-4}
				\textbf{Simulation setting} & \textbf{Average (sd) of $\hat\pi_1$}  & \textbf{Average (sd) of $|\hat\muv_1 - \muv_1|_{L_1}$}& \textbf{Average (sd) of $|\hat\Sigmav_1 - \Sigmav_1|_{L_1}$}\\
				\midrule
				K=5, n=100	&	0.4996 (0.009)	&	0.3789 (0.1535)	&	1.281 (0.3175)	\\
				K=5, n=200	&	0.5005 (0.0038)	&	0.2425 (0.1021)	&	0.8694 (0.2536)	\\
				K=5, n=500	&	0.4999 (0.0023)	&	0.1489 (0.0592)	&	0.574 (0.1597)	\\
				\midrule							
				K=10, n=100	&	0.4896 (0.0129)	&	0.9975 (0.2892)	&	10.1231 (1.8573)\\
				K=10, n=200	&	0.496 (0.0067)	&	0.7483 (0.2515)	&	7.1662 (1.2323)	\\
				K=10, n=500	&	0.5005 (0.0024)	&	0.588 (0.1657)	&	7.8635 (2.0106)	\\
				\midrule							
				K=20, n=100	&	0.4546 (0.0212)	&	1.7992 (0.3343)	&	26.4108 (4.9824)\\
				K=20, n=200	&	0.4991 (0.0047)	&	1.111 (0.266)	&	15.718 (1.5242)	\\
				K=20, n=500	&	0.4998 (6e-04)	&	0.7108 (0.1562)	&	10.1909 (0.9694)\\
				
				\midrule
				\midrule
				&\multicolumn{3}{@{}c@{}}{\textbf{ Component 2 ($\pi_2 = 0.5$)}}\\
				\cmidrule{2-4}
				\textbf{Simulation setting} & \textbf{Average (sd) of $\hat\pi_2$}  & \textbf{Average (sd) of $|\hat\muv_2 - \muv_2|_{L_1}$}& \textbf{Average (sd) of $|\hat\Sigmav_2 - \Sigmav_2|_{L_1}$}  \\
				\midrule
				K=5, n=100	&	0.5004 (0.009)	&	0.3094 (0.1344)	&	0.9111 (0.2978)	\\
				K=5, n=200	&	0.4995 (0.0038)	&	0.203 (0.0812)	&	0.6462 (0.2292)	\\
				K=5, n=500	&	0.5001 (0.0023)	&	0.1293 (0.0604)	&	0.3882 (0.1385)	\\
				\midrule							
				K=10, n=100	&	0.5104 (0.0129)	&	0.59 (0.1926)	&	3.5965 (2.1616)	\\
				K=10, n=200	&	0.504 (0.0067)	&	0.3837 (0.1212)	&	2.0346 (0.9415)	\\
				K=10, n=500	&	0.4995 (0.0024)	&	0.5144 (0.1382)	&	5.9088 (1.0227)	\\
				\midrule							
				K=20, n=100	&	0.5454 (0.0212)	&	2.3534 (0.6456)	&	43.53 (12.8879)	\\
				K=20, n=200	&	0.5009 (0.0047)	&	0.8151 (0.2012)	&	9.4582 (2.157)	\\
				K=20, n=500	&	0.5002 (6e-04)	&	0.5151 (0.1226)	&	5.7646 (0.5573)	\\
				
				\bottomrule
		\end{tabular*}}
	\end{table}
\end{center}

\begin{table}[!ht]
\caption{Cross tabulation of the clusters obtained by our proposed algorithm and Dirichlet-multinomial mixture (DMM) model  on all four real datasets.\label{ctRD1}}
\textbf{The \texttt{Ferretti} gut microbiome subset}
\begin{center}
\begin{tabular*}{0.8\textwidth}{@{\extracolsep{\fill}}lcclcc}
\hline
& \multicolumn{2}{c}{Proposed algorithm (ARI: 0.81)}&\multicolumn{3}{c}{DMM model (ARI: 0.73)}\\\cline{2-3}\cline{5-6}
& \multicolumn{2}{c}{Estimated Clusters}&\multicolumn{3}{c}{Estimated Clusters}\\
&1&2&&1&2\\
\hline
Infant&18 & 1 & &17&2\\
Adult & 1 & 22& & 1&22\\
\hline
\end{tabular*}
\end{center}
\textbf{The \texttt{Ferretti} Oral microbiome subset}
\begin{center}
\begin{tabular*}{0.8\textwidth}{@{\extracolsep{\fill}}lcclcc}
\hline
& \multicolumn{2}{c}{Proposed algorithm (ARI: 1)}&\multicolumn{3}{c}{DMM model (ARI: 1)}\\\cline{2-3}\cline{5-6}
& \multicolumn{2}{c}{Estimated Clusters}&\multicolumn{3}{c}{Estimated Clusters}\\
&1&2&&1&2\\
\hline
Infant&23 & - & &23&-\\
Adult & - & 23& & -&23\\
\hline
\end{tabular*}
\end{center}
\textbf{The \texttt{Shi} dataset}
\begin{center}
\begin{tabular*}{0.8\textwidth}{@{\extracolsep{\fill}}lcclcc}
\hline
& \multicolumn{2}{c}{Proposed algorithm (ARI: 0.49)}&\multicolumn{3}{c}{DMM model (ARI: 0.43)}\\\cline{2-3}\cline{5-6}
& \multicolumn{2}{c}{Estimated Clusters}&\multicolumn{3}{c}{Estimated Clusters}\\
&1&2&&1&2\\
\hline
Periodontitis&18 & 6& &18&6\\
Recovered& 1 & 23& & 2&22\\
\hline
\end{tabular*}
\end{center}
\textbf{The \texttt{atlas1006} dataset}
\begin{center}
\begin{tabular*}{0.8\textwidth}{@{\extracolsep{\fill}}lcccclcccc}
\hline
& \multicolumn{3}{c}{Proposed algorithm (ARI: 0.048)}&\multicolumn{5}{c}{DMM model (ARI: 0.029)}\\\cline{2-4}\cline{6-9}
& \multicolumn{3}{c}{Estimated Clusters}&\multicolumn{5}{c}{Estimated Clusters}\\
&1&2&3&&1&2&3&4\\
\hline
Lean&103 & 118&284 &&176&104&136&89\\
Overweight& 51 & 29& 117&&66&47&55&29\\
Obese&147&98&98&&57&133&59&94\\
\hline
\end{tabular*}
\end{center}
\end{table}

\clearpage
\appendix
\section{Mathematical Detail}\label{math_detail}
Consider the following transformed parameter $\etav$ from $\Yv$:
\begin{equation*}
\etav = B\Yv, \quad \text{ where }B=\begin{pmatrix}
1&0&\dots&0\\
0&1&\dots&0\\
\vdots&\vdots&\dots&\vdots\\
0&0&\dots&1\\
0&0&\dots&0
\end{pmatrix},
\end{equation*}
is a $(K+1)\times K$ matrix which takes the form as an identity matrix attached by a row of K zeros. Given the assumption that the true distribution of $\Yv$ is $\normal(\muv,\Sigmav)$, we have the true distribution of $\etav$ to be Gaussian too, with mean $\tilde{\muv}$ and covariance matrix $\tilde{\Sigmav}$, where
\begin{equation*}
\setstretch{1.5}
\tilde{\muv} = B\muv = (\muv,0)^\top; \tilde{\Sigmav} = B\Sigmav B^\top = \begin{pmatrix}
\Sigmav&\vline &\0v_{K\times 1}\\
\hline
\0v_{1\times K}&\vline&0
\end{pmatrix}.
\end{equation*}

For computational convenience, we further assume that $V$ has a diagonal structure, with each diagonal element denoted as $v_k^2$ such that
\begin{equation*}
v_k^2 = \begin{cases}
v_k^2,&k=1,\dots,K\\
0,&k=K+1.
\end{cases}
\end{equation*} We also denote the $k-$th element of $\mv$ as $m_k$ such that
\begin{equation*}
m_k = \begin{cases}
m_k, &k=1,\dots,K\\
0,&k=K+1.
\end{cases}
\end{equation*}
\par 
Recall that we have the following decomposition of the ELBO
\begin{equation*}
F(q(\etav),\wv) = F(\mv,V) = - \int q(\etav)\log q(\etav) d\etav + \int q(\etav)\log p(\etav) d\etav + \int q(\etav)\log p(\wv|\etav) d\etav;
\end{equation*}
among which, the first integral by definition is the entropy of the variational Gaussian distribution $q(\etav|\mv,V)$:
\begin{equation*}
- \int q(\etav)\log q(\etav) d\etav =-\E_{q(\etav|\mv,V)}\left(q(\etav)\right) = \half\sum_{k=1}^K{\log(v_k^2)}+\dfrac{K}{2}\log(2\pi)+\dfrac{K}{2}.
\end{equation*}
The second integral can be evaluated explicitly as well, which turn into the expected value of the log density function of $p(\etav) = \normal(\etav|\tilde \muv,\tilde \Sigmav)$ with respect to $q(\etav|\mv,V)$:
\begin{equation*}
\begin{split}
\int q(\etav)\log p(\etav) d\etav &= \E_{q(\etav|\mv,V)}\left(\log p(\etav)\right)\\
&=\E_{q(\etav|\mv,V)}\left(-\dfrac{K}{2}\log(2\pi)-\half\log|\tilde \Sigmav|-\half(\etav-\muv)^\top\tilde\Sigmav^{-1}(\etav-\muv)\right)\\
&=-\dfrac{K}{2}\log(2\pi)-\half\log|\tilde \Sigmav|-\half(\mv-\tilde \muv)^\top\tilde\Sigmav^{-1}(\mv-\tilde\muv)-\half \Tr(\tilde\Sigmav^{-1}V).
\end{split}
\end{equation*}
Due to the special structure of $\tilde \Sigmav$, we have $|\tilde \Sigmav| = 0$ and $\tilde \Sigmav^{-1}$ does not exist, which brings in a computational issue. Therefore, we substitute $|\tilde \Sigmav|$ by $|\Sigmav|=|B^\top\tilde\Sigmav B|$ and $\tilde \Sigmav^{-1}$ by the generalized inverse of $\tilde \Sigmav$
\begin{equation*}
\tilde \Sigmav^{\ast} = \begin{pmatrix}
\Sigmav^{-1}&\vline &\0v_{K\times 1}\\
\hline
\0v_{1\times K}&\vline&0
\end{pmatrix}.
\end{equation*}
Hence we have 
\begin{equation*}
\begin{split}
\int q(\etav)\log p(\etav) d\etav &=-\dfrac{K}{2}\log(2\pi)-\half\log|\Sigmav|-\half(\mv-\tilde \muv)^\top\tilde\Sigmav^{\ast}(\mv-\tilde\muv)-\half \Tr(\tilde\Sigmav^{\ast}V)\\
&=-\dfrac{K}{2}\log(2\pi)-\half\log|B^\top\tilde\Sigmav B|-\half(\mv-\tilde \muv)^\top\tilde\Sigmav^{\ast}(\mv-\tilde\muv)-\half \Tr(\tilde\Sigmav^{\ast}V).
\end{split}
\end{equation*}
The third integral is intractable, because of the log-sum exponential term. We upper bound this term with a Taylor expansion similar to previous literature \cite{blei2006} resulting in the following
\begin{equation*}
\begin{split}
\E_{q(\etav|\mv,V)}\left[\log\left(\sum_{k=1}^{K+1}{\exp\eta_k}\right)\right] &\leq \xi^{-1}\left\{\sum_{k=1}^{K+1}\E_{q(\etav|\mv,V)}\left[\exp(\eta_k)\right]\right\}-1+\log(\xi)\\
&=\xi^{-1}\left\{\sum_{k=1}^{K+1} \exp\left(m_k+\dfrac{v_k^2}{2}\right)\right\}-1+\log(\xi).
\end{split}
\end{equation*}
Therefore, the third integral is lower bounded by 
\begin{equation*}
\begin{split}
\int q(\etav)\log p(\wv|\etav) d\etav&= \E_{q(\etav|\mv,V)}\left[\wv^\top\etav-\sum_{k=1}^{K+1}{w_k\log\left(\sum_{k=1}^{K+1}{\exp\eta_k}\right)}\right]\\
&= \wv^\top\mv - \left(\sum_{k=1}^{K+1} w_k\right)\E_{q(\etav|\mv,V)}\left[\log\left(\sum_{k=1}^{K+1}{\exp\eta_k}\right)\right]\\
&\geq \wv^\top\mv- \left(\sum_{k=1}^{K+1} w_k\right)\left\{\xi^{-1}\left[\sum_{k=1}^{K+1}\exp\left(m_k+\dfrac{v_k^2}{2}\right)\right]-1+\log(\xi)\right\}.
\end{split}
\end{equation*}
Combining all three integrals, we obtain a concave variational Gaussian lower bound to the model evidence
\begin{equation*}
\setstretch{2}
\begin{split}
\tilde{F}\left(\mv,V,\tilde{\muv},\tilde{\Sigmav},\xi\right) &= \wv^\top\mv - \left(\sum_{k=1}^{K+1} w_k\right)\left\{\xi^{-1}\left[\sum_{k=1}^{K+1}\exp\left(m_k+\dfrac{v_k^2}{2}\right)\right]-1+\log(\xi)\right\}\\
&-\half\log|B^\top\tilde\Sigmav B|-\half(\mv-\tilde \muv)^\top\tilde\Sigmav^{\ast}(\mv-\tilde \muv)-\half \Tr(\tilde\Sigmav^{\ast}V)\\
&+\half\sum_{k=1}^K\log(v_k^2)+\dfrac{K}{2}.
\end{split}
\end{equation*}
We maximize this lower bound with respect to the variational parameters $\xi,\mv,V$.\par 
First, we maximize the lower bound~\ref{vglb} with respect to $\xi$. The derivative with respect to $\xi$ is 
\begin{equation*}
\dfrac{\partial\tilde F}{\partial \xi} = \left(\sum_{k=1}^{K+1}w_k\right)\left\{-\xi^{-2}\left[\sum_{k=1}^{K+1}\exp\left(m_k+\dfrac{v_k^2}{2}\right)\right]+\xi^{-1}\right\},
\end{equation*}
which yields an optimizer at
\begin{equation*}
\hat \xi = \sum_{k=1}^{K+1}\exp\left(m_k+\dfrac{v_k^2}{2}\right).
\end{equation*}
Second, we maximize with respect to $\mv$, of which the derivative is given as
\begin{equation*}
\dfrac{\partial \tilde{F}}{\partial \mv} = \wv - \tilde{\Sigmav}^{\ast}(\mv-\tilde \muv)-\left(\sum_{k=1}^{K+1} \wv_k\right)\xi^{-1}\exp\left(\mv+\dfrac{\vv^2}{2}\right),
\end{equation*}
with $\vv^2=(v_1^2,\dots,v_K^2,0)$ denoting the diagonal element of $V$ as a vector. There is no analytical solution to this derivative and so we use Newton's method to approximate the root to this derivative, with a constrain that the $(K+1)-$th element is zero. The procedure requires the Hessian matrix with respect to $\mv$:
\begin{equation*}
H_{\mv} = -\tilde\Sigmav^{\ast} - \left(\sum_{k=1}^{K+1} \wv_k\right)\xi^{-1}\diag \left\{\exp\left(\mv+\dfrac{\vv^2}{2}\right)\right\}.
\end{equation*}
Finally, we optimize with respect to $v_k$, for $k=1,\dots,K$ and always set $v_{K+1}$ as zero. Again, there are no analytical solutions and Newton's method is used for each coordinate. The first and second derivatives with respect to $v_k$ for $k=1,\dots,K$ are given as follows
\begin{equation*}
\begin{split}
\dfrac{\partial \tilde{F}}{\partial v_k} &= v_k^{-1}-v_k\tilde \Sigmav_{k,k}^{\ast} - \left(\sum_{k=1}^{K+1} \wv_k\right)\xi^{-1}\exp\left(m_k+\dfrac{v_k^2}{2}\right)v_k;\\
\dfrac{\partial^2 \tilde{F}}{\partial v_k^2} &= -v_k^{-2}-\left(\sum_{k=1}^{K+1} \wv_k\right)\xi^{-1}\exp\left(m_k+\dfrac{v_k^2}{2}\right)\left(v_k^2 + 1\right).
\end{split}	
\end{equation*}
At each iteration of the variational EM algorithm, when we maximize the variational Gaussian lower bound $\tilde{F}(\mv,V,\tilde{\muv},\tilde{\Sigmav},\xi)$ with respect to the variational parameter set $(\xi,\mv,V)$, we take one step of update based on the optimization discussed above.

\end{document}